\documentclass[prologue,dvipsnames,sigplan,screen]{acmart}

\def\fullversion

\setcopyright{acmcopyright}
\copyrightyear{2018}
\acmYear{2018}
\acmDOI{XXXXXXX.XXXXXXX}

\usepackage{booktabs}   
\usepackage{subcaption} 
\usepackage{caption}
\usepackage{colortbl}
\usepackage{makecell}
\usepackage{bigstrut}
\usepackage{microtype}
\usepackage{tabularx}
\usepackage{multirow}
\usepackage{multicol}
\usepackage{booktabs}
\usepackage{rotating}
\usepackage{lipsum}
\usepackage{balance}
\usepackage{mfirstuc}
\usepackage{titlecaps}
\usepackage{listings}
\usepackage{float}
\usepackage{arydshln}
\usepackage[rightcaption]{sidecap}
\usepackage{setspace}
\usepackage{tcolorbox}
\usepackage[normalem]{ulem}
\useunder{\uline}{\ul}{}

\usepackage{xcolor}
\usepackage[font=normalfont,labelfont=bf,skip=2pt]{caption}

\newcommand{\oref}[1]{\hyperref[#1]{#1}}

\usepackage{xr}
\usepackage{etoolbox}
\newcommand{\ifconference}[1]{{{\ifx\fullversion\undefined{#1}\fi}\xspace}}
\newcommand{\iffullversion}[1]{{{\ifx\conference\undefined{#1}\fi}\xspace}}

\usepackage{graphicx}  
\newcommand{\hide}[1]{} 
\usepackage{xspace}
\usepackage{textcomp}
\usepackage{comment} 
\usepackage{verbatim}
\usepackage{url}

\usepackage{pifont}

\usepackage[shortlabels]{enumitem}

\setlist{topsep=0.3em,itemsep=0.2em,parsep=0.1em,leftmargin=*}

\usepackage{float}
\usepackage[font={small},aboveskip=0em, belowskip=1em]{caption}

\usepackage[labelfont=bf,list=true,skip=0em]{subcaption}
\usepackage[rightcaption]{sidecap}

\usepackage{wrapfig}

\usepackage{array}
\newcolumntype{L}[1]{>{\raggedright\let\newline\\\arraybackslash\hspace{0pt}}m{#1}}
\newcolumntype{C}[1]{>{\centering\let\newline\\\arraybackslash\hspace{0pt}}m{#1}}
\newcolumntype{R}[1]{>{\raggedleft\let\newline\\\arraybackslash\hspace{0pt}}m{#1}}
\newcolumntype{B}{>{\bf}c}
\usepackage{rotating}

\usepackage{booktabs} 
\usepackage{multicol,multirow}
\usepackage{longtable} 
\usepackage{supertabular} 
\usepackage{colortbl}
\usepackage{bigstrut}

\newcommand{\myparagraph}[1]{\vspace{.03in}\noindent {\boldmath\bf #1\unboldmath}}
\newcommand{\rmv}[1]{\textcolor{red}{\sout{}}}

\usepackage[ruled,lined,linesnumbered,noend]{algorithm2e}
\usepackage[noend]{algpseudocode}

\makeatletter
\patchcmd{\@algocf@start}
{-1.5em}
{0pt}
{}{}
\newcommand{\nosemic}{\renewcommand{\@endalgocfline}{\relax}}
\newcommand{\dosemic}{\renewcommand{\@endalgocfline}{\algocf@endline}}

\SetSideCommentLeft
\SetKwInput{notations}{Notations}
\SetKwInput{notes}{Notes}
\SetKwInput{maintains}{Maintains}

\SetKwProg{myfunc}{Function}{}{}
\SetKwFor{parForEach}{ParallelForEach}{do}{endfor}
\SetKwFor{Justrepeat}{Repeat}{}{}

\definecolor{dpcol}{RGB}{0,160,240}

\newcommand{\yihan}[1]{{\color{blue}{\bf Yihan:} #1}}

\newcommand{\helen}[1]{{\color{purple}{\bf Helen:} #1}}

\SetCommentSty{mycommfont}

\usepackage{mdframed}
\definecolor{framelinecolor}{RGB}{68,114,196}
\mdfdefinestyle{mystyle}{linecolor=framelinecolor,innertopmargin=1pt,innerbottommargin=2pt,backgroundcolor=gray!20,skipabove=2pt,skipbelow=0pt}
\mdfdefinestyle{densestyle}{linecolor=framelinecolor,innertopmargin=0,innerbottommargin=0,leftmargin=0,rightmargin=0,backgroundcolor=gray!20}
\mdfdefinestyle{compactcode}{linecolor=framelinecolor,innertopmargin=1pt,innerbottommargin=1pt,backgroundcolor=gray!20,skipabove=0pt,skipbelow=0pt,leftmargin=0,rightmargin=0}

\usepackage{framed}

\usepackage{listings}

\newdimen\zzsize
\newdimen\kwsize
\newcommand{\basicstyle}{\fontsize{\zzsize}{1\zzsize}\ttfamily}
\newcommand{\keywordstyle}{\fontsize{\kwsize}{1\kwsize}\ttfamily\bf}

\newdimen\zzlstwidth
\usepackage{cleveref}
\crefname{appendix}{Appendix}{Appendix}
\crefname{section}{Sec.}{Secs.}
\crefname{theorem}{Thm.}{Thm.}
\crefname{lemma}{Lem.}{Lem.}
\crefname{corollary}{Col.}{Col.}
\crefname{table}{Tab.}{Tab.}
\crefname{algorithm}{Alg.}{Alg.}
\crefname{figure}{Fig.}{Fig.}
\crefname{fact}{Fact}{Fact}
\Crefname{table}{Tab.}{Tab.}
\crefname{problem}{Problem}{Problem}

\usepackage{tikz} 

\newcommand*{\rom}[1]{\expandafter\@slowromancap\romannumeral #1@}

\newcommand{\cpam}{\textsf{CPAM}\xspace}

\newcommand{\pam}{\textsf{PAM}\xspace}
\newcommand{\pma}{\textsf{PMA}\xspace}

\newcommand{\masstree}{\textsf{Mass-tree}\xspace}
\newcommand{\bwtree}{\textsf{Bw-tree}\xspace}
\newcommand{\ourtree}{\mbox{\textsf{PaB-tree}}\xspace}

\newcommand{\ourlib}{\textsf{PaB-tree}\xspace}

\newcommand{\pbtree}{\textsf{PaB-tree}\xspace}

\newcommand{\ske}{\texttt{Skew}\xspace}
\newcommand{\adv}{\texttt{Adv}\xspace}
\newcommand{\uniform}{\texttt{Uni}\xspace}

\newcommand{\find}{\textsc{Find}\xspace}
\newcommand{\rreport}{\textsc{Range-Report}\xspace}
\newcommand{\neighbor}{\textsc{Neighbor}\xspace}
\newcommand{\flatten}{\textsc{Flatten}\xspace}

\newcommand{\osm}{\texttt{OSM}\xspace}
\newcommand{\planet}{\texttt{PLANET}\xspace}
\newcommand{\redd}{\texttt{RED}\xspace}
\newcommand{\reference}{\texttt{REF}\xspace}
\newcommand{\sem}{\texttt{SEM}\xspace}

\newcommand{\separator}{separator}
\newcommand{\heavynode}{heavy node}
\newcommand{\midb}{\mathit{2B}}
\newcommand{\overflow}{\textsc{Promote}}
\newcommand{\size}{\mathit{size}}
\newcommand{\sep}{separator}
\newcommand{\nd}{subtree}
\newcommand{\seps}{\mathit{separators}}
\newcommand{\nds}{\mathit{subtrees}}

\newcommand{\batchdelete}{\textsc{BatchDelete}}

\SetKw{MIN}{min}
\SetKw{MAX}{max}
\SetKw{OR}{or}
\SetKw{AND}{and}

\newcommand{\codeskip}{{\vspace{.05in}}}

\newcommand{\defn}[1]{\emph{\textbf{#1}}} 

\newcommand{\puf}{H}
\newcommand{\newpuf}{H'}
\newcommand{\pufseq}{\mathcal{H}}
\newcommand{\nseq}{\mathcal{N}}
\newcommand{\mathtext}[1]{{\mathit{#1}}}
\newcommand{\ub}{\mathtext{4B}\xspace}
\newcommand{\lb}{\mathtext{B}\xspace}
\newcommand{\nodes}{\mathtext{nodes}\xspace}
\newcommand{\splits}{\mathtext{splits}\xspace}

\newcommand{\bt}{$B^+$-tree\xspace}

\newcommand{\btree}{\bt}

\newcommand{\batchinsert}{\textsc{BatchInsert}}

\newcommand{\bbplustree}{B/B$^+$-tree}
\newcommand{\bplustree}{B$^+$-tree}

\renewcommand{\defn}[1]       {{\textit{\textbf{\boldmath #1}}}}

\newcommand{\tabref}[1]        {Table~\ref{tab:#1}}
\newcommand{\secref}[1]         {Section~\ref{sec:#1}}

\newtheorem{theorem}{Theorem}[section]
\newtheorem{lemma}[theorem]{Lemma}
\newtheorem{corollary}[theorem]{Corollary}

\newtheorem{fact}[theorem]{Fact}

\let \originalleft \left
\let\originalright\right
\renewcommand{\left}{\mathopen{}\mathclose\bgroup\originalleft}
\renewcommand{\right}{\aftergroup\egroup\originalright}

\usepackage{scalerel} 

\newtheoremstyle{exampstyle}
{.5em} 
{1em} 
{\it} 
{.5em} 
{\it \bfseries} 
{.} 
{.5em} 
{} 

\makeatletter
\renewenvironment{proof}[1][\proofname]{\par
  \vspace{-2\topsep}
  \pushQED{\qed}%
  \normalfont
  \topsep0pt \partopsep0pt 
  \trivlist
  \item[\hskip\labelsep
              \itshape
              #1\@addpunct{.}]\ignorespaces
}{%
  \popQED\endtrivlist\@endpefalse
}

\crefname{table}{Tab.}{Tab.}
\crefname{figure}{Fig.}{Fig.}

\makeatletter
\newcommand{\pushright}[1]{\ifmeasuring@#1\else\omit\hfill$\displaystyle#1$\fi\ignorespaces}
\newcommand{\pushleft}[1]{\ifmeasuring@#1\else\omit$\displaystyle#1$\hfill\fi\ignorespaces}
\makeatother

\begin{document}

\makeatletter
\gdef\@copyrightpermission{
	\begin{minipage}{0.2\columnwidth}
		\href{https://creativecommons.org/licenses/by/4.0/}{\includegraphics[width=0.90\textwidth]{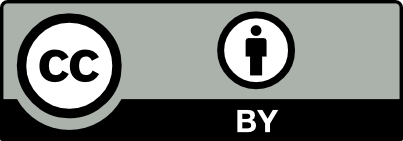}}
	\end{minipage}\hfill
	\begin{minipage}{0.8\columnwidth}
		\href{https://creativecommons.org/licenses/by/4.0/}{This work is licensed under a Creative Commons Attribution International 4.0 License.}
	\end{minipage}
	\vspace{5pt}
}
\makeatother


\ifconference{ \title{\ourtree{}: Parallel \btree{} with Batch Updates } }
\iffullversion{ \title{\ourtree{}: Parallel \btree{} with Batch Updates } }


\settopmatter{authorsperrow=4}
\author{Ziyang Men}
\email{ziyang.men@email.ucr.edu}
\affiliation{\institution{University of California, Riverside}\country{}}
\author{Yan Gu}
\email{ygu@cs.ucr.edu}
\affiliation{\institution{University of California, Riverside}\country{}}
\author{Yihan Sun}
\email{yihans@cs.ucr.edu}
\affiliation{\institution{University of California, Riverside}\country{}}
\author{Helen Xu}
\email{hxu615@gatech.edu}
\affiliation{\institution{Georgia Institute of Technology}\country{}}


\begin{abstract}
  The \btree{} is a ubiquitous search index due to its provable
  I/O (cache) efficiency. Yet unlike batch-parallel binary search trees, which are both
  fast and widely available, no batch-parallel \btree{} offers strong empirical
  performance together with provable, near-optimal I/O efficiency. Prior
  \btree{}-based designs are either parallel but purely theoretical, with no
  high-performance implementation, or concurrent indexes such as the Bw-tree and
  Mass-tree, which run an order of magnitude slower on batch updates.

  We present \ourtree{}, the first batch-parallel \btree{} with guarantees on
  work, span, and cache complexity alongside strong practical performance. Its
  core challenge---repairing node overflows and underflows without eroding the
  tree's I/O advantage---is resolved by one simple invariant: a repair never
  touches a sibling and is always pushed to the parent. A \textsc{Promote}
  primitive reduces batch construction and insertion to repeated node promotion,
  and batch deletion to reinsertion. Batch updates take $O(B\log_B n)$ work and
  $O(\log_B n)$ I/O per element, matching the sequential \btree{} bounds, with
  polylogarithmic span. 
  Against highly optimized parallel and concurrent baselines, \ourtree{}
  delivers the best overall performance trade-off: up to $1.4\times$ faster on
  construction and up to $2.0\times$ faster on batch insertion than the
  strongest baseline, while preserving \btree{} query efficiency.
\end{abstract}

\ccsdesc[500]{Computer systems organization~Embedded systems}
\ccsdesc[300]{Computer systems organization~Redundancy}
\ccsdesc{Computer systems organization~Robotics}
\ccsdesc[100]{Networks~Network reliability}


\renewcommand\footnotetextcopyrightpermission[1]{} 

\maketitle
\section{Introduction}\label{sec:intro}

\bbplustree{s}~\cite{bayer1972, comer1979} are among the most fundamental data
structures and search indexes, widely adopted in both in-memory and
external-memory systems.  In this paper, we focus on the in-memory setting.
Unlike binary search trees (BSTs), which use a single key as a separator,
\bbplustree{s} keep a $\Theta(B)$-way fan-out so that each node aligns with the
hardware access unit (typically a few cache lines in the in-memory setting).
Every node access therefore fetches $\Theta(B)$ keys for comparison, and a
single access is logically equivalent to descending $\Theta(\log B)$ levels of a
BST.  This design positions the \bbplustree{} as one of the most canonical data
structures for achieving \defn{provable I/O (cache) efficiency}.  Backed by this
theoretical guarantee, in-memory \bbplustree{s} typically outperform binary
trees by 2--8$\times$ in the sequential setting, and therefore have been
natively offered across a wide spectrum of software
libraries~\cite{abseil2018btree,
  bingmann2018tlx,bingmann2013stx,rust2024btreemap,google2024gobtree}.

\hide{In this paper, we focus on efficient \emph{parallel in-memory B-tree
    design and implementation}. As shown in Table~\ref{...}, our proposed
  parallel B-tree maintains the cache-friendliness of the sequential B-tree
  without sacrificing parallelism compared to BSTs. \helen{measure cache misses
    in pb-tree vs BSTs?}  }

\myparagraph{Batch-Parallel Paradigm.}  Driven by growing data scales and the
multicore era, batch-parallel data structures have attracted significant
research
interest~\cite{sun2018pam,blelloch2016just,wheatman2024byo,wheatman2021parallel,dhulipala2022pac,men2025parallel,men2026parallel,dong2021efficient,pandey2021terrace,WheatmanXu18,bpcsr,WheatmanXu21,shun2014phase,ptreedb,wang2020theoretically,wang2022pargeo,blelloch2016parallel,kang2023skew,zhao2026pimzdtree,acar2019parallel,anderson2020work,kang2021processing,kang2023pim,dhulipala2019low,tseng2019batch}.
These structures apply updates (insertions or deletions) in batches in parallel,
and have been adopted in graph
processing~\cite{dhulipala2019low,wheatman2024byo,fineman2019nearly,wang2023fast,dong2021efficient,pandey2021terrace},
databases~\cite{ptreedb,sewall2011palm}, parallel algorithms for dynamic
programming~\cite{ding2024parallel,gu2023parallel}, computational
geometry~\cite{blelloch2022parallel,men2026parallel}, vector similarity
search~\cite{men2025parallel, zhao2025optimal, zhao2026pimzdtree}, etc.

Although both \bbplustree{s}~\cite{bayer1972, comer1979} and
BSTs~\cite{adelsonvelskii1963algorithm, guibas1978Dichromatic,
  nievergelt1973binary} were introduced over 50 years ago, and while
\bbplustree{s} are the go-to index design due to their cache-friendliness, there
is a striking disparity in the maturity of their batch-parallel
counterparts. Parallel BST implementations have been studied extensively and are
widely accessible in practice, with examples including the
PAM~\cite{sun2018pam,blelloch2016just}, Aspen~\cite{dhulipala2019low},
CPAM~\cite{dhulipala2022pac}, and MCSTL~\cite{frias2007parallelization}.  Many
of them are also theoretically efficient, achieving high parallelism while
preserving optimal or near-optimal sequential work bounds.  In contrast, there
is a notable lack of accessible batch-parallel \bbplustree{} implementations
with both strong empirical performance and provable, near-optimal I/O
efficiency.  Existing efforts are either purely theoretical~\cite{PVW83,
  higham1994maintaining, goodrich2025parallel}, with no high-performance
implementation, or concurrent, such as the well-known Bw-tree~\cite{wang2018building} and the Mass-tree~\cite{mao2012cache}. However,
concurrent structures do not exploit the inherent advantages of batching;
consistent with prior
findings~\cite{sun2018pam},~\cref{sec:exp} 
shows that they run an order of magnitude slower than parallel trees in batch
updates.  Other studies have explored binary trees with expanded node
sizes~\cite{dhulipala2019low,dhulipala2022pac}, but they remain a suboptimal
compromise for I/O efficiency (also see \cref{fig:intro}), due to looser
tree-height bounds compared to a true \bbplustree{}, and none offer non-trivial
cache complexity.

\myparagraph{Challenges in Designing Efficient Parallel \bbplustree{s}.}  The
primary challenge in parallelizing \bbplustree{s} lies in achieving low-cost
structural repairs without eroding their I/O advantage.  While the high fan-out
in a \bbplustree{} decreases memory transfers (i.e., the cache complexity), it
does not reduce the total work (in terms of the number of comparisons), and is
inherently more complicated to deal with.  Whereas BST rotations remain strictly
local, the high fan-out nature of \bbplustree{s} may cause structural imbalances
(overflows or underflows) to propagate across neighboring parallel subtasks,
demanding dedicated efforts to repair.  Since the reduction in memory access,
$\Theta(\log_2 B)$, is relatively modest in practice, any overhead added to
reconcile structural modifications is subtracted from the sole term that
constitutes the benefit.  Consequently, a merely correct design is easily slower
than a well-tuned parallel BST.

This raises a critical question: \textit{Can the
high fan-out design of \bbplustree{s} still offer provable cache efficiency in
the parallel setting, and can this advantage translate into better performance
in practice?}  In this paper, we answer the
question affirmatively by presenting the \pbtree{}, the first batch-parallel
\bplustree{} implementation with both theoretical guarantees (work, span, cache
complexity) and strong practical performance across a wide range of workloads.

\myparagraph{Approach Overview.} In light of the challenges discussed above, our
design prioritizes conceptual simplicity to minimize parallel overhead during
structural repairs.  The design follows a single rule: \emph{a repair never
  touches a sibling; it is always pushed to the parent.}

The difficulty is that a repair must never reach sideways to a sibling, which
another parallel task may be modifying at the same time; instead, every repair
travels up the ancestor path the recursion already owns. We realize this by
having each recursive call return a \emph{heavy node}—a possibly oversized node
that the parent absorbs and re-partitions—rather than a fixed-shape subtree, and
by keeping a node-capacity margin that lets the more involved deletion cost
amortize fully.

A single overflow-like operation (our \overflow{} primitive detailed
in~\cref{sec:algorithms}) resolves an oversized heavy node, and both construction
and insertion reduce to repeated applications of it.  Underflow in batch
deletion, however, is more challenging to handle.  Our solution reduces
underflow repair to insertion---an operation we have already made parallel and
cache-efficient---so no node ever consults a sibling.  Specifically, given a set
of underflowed subtrees invalidated by a batch deletion, we carefully identify a
few entry points for them to be reinserted, letting them share the work, such
that the total overhead can be bounded in an amortized fashion
(\cref{thm:batch_delete}). \cref{fig:osm} shows that the reinserted volume
also stays modest in practice, so the reduction is inexpensive as well as
simple.

\begin{figure}[t]
	\includegraphics[width=0.48\textwidth]{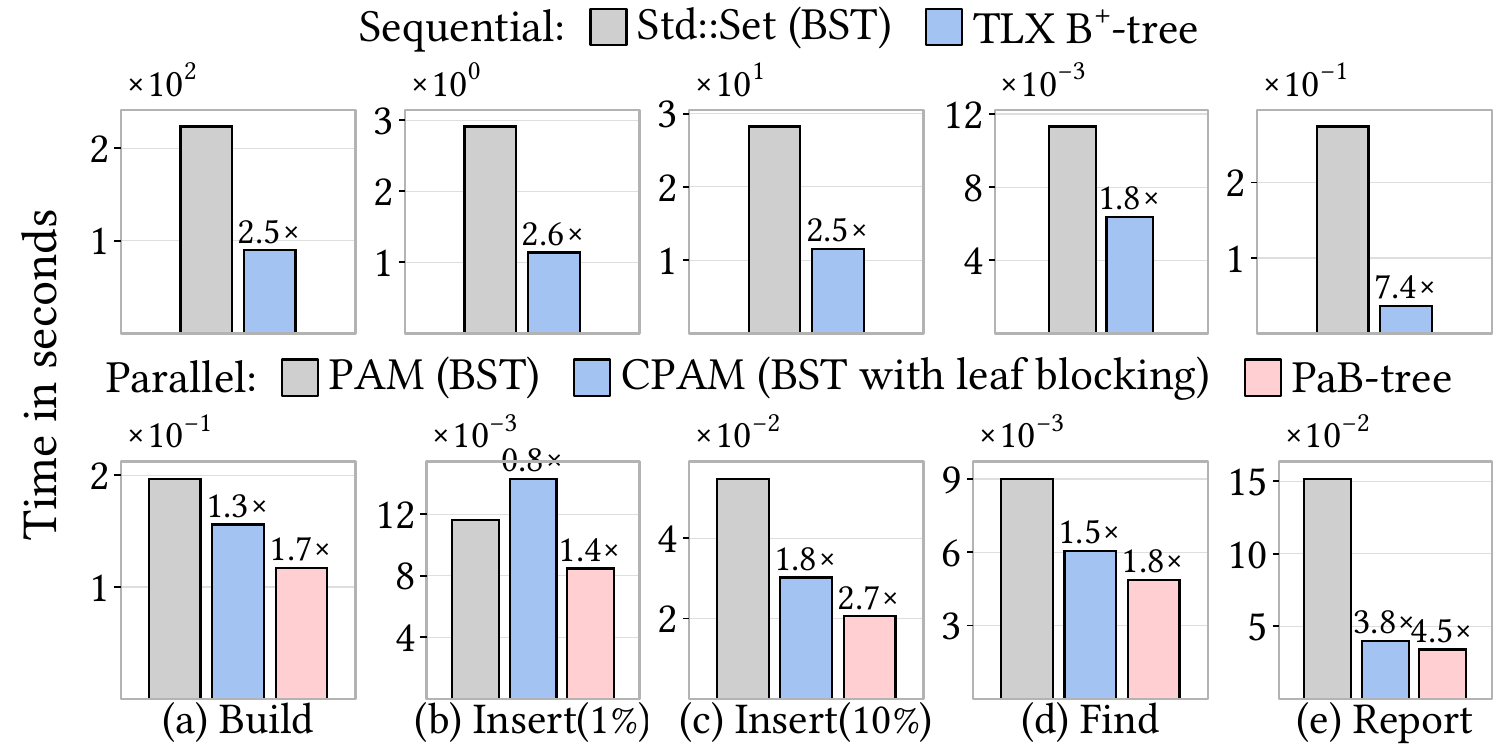}
	\caption{\textbf{Running time (in seconds) of the sequential and parallel
            BSTs and \btree{s}. Lower is better.}  We measure the time under a
          uniform distribution, including the time to build a tree containing $10^8$
          entries and the time to insert 1\% and 10\% of entries into the tree.
          The query time is measured by searching $10^6$ entries, and reporting
          $[10^3, 10^5]$ entries in a range.}
	\label{fig:intro}
\end{figure}


\myparagraph{Results Summary.}  For both batch insertion and deletion, \pbtree{}
achieves provable bounds on work, span, and cache complexity.  A batch insertion or
deletion incurs $O(B\log_B n)$ work and $O(\log_B n)$ I/O cost per element,
matching the bounds of a sequential \btree, and both operations achieve polylogarithmic span (\cref{thm:btree_insertion} and \cref{thm:batch_delete}).  The theoretical efficiency also translates into strong
performance in practice. \cref{fig:intro} confirms this observation: \pbtree{} matches the cache-friendliness of sequential
\btree without sacrificing parallelism, and excels when processing large batches.

\cref{fig:summary} confirms this trend on a more comprehensive benchmark across
different operations, data distributions, and more baselines.  Overall,
\pbtree{} achieves the best performance trade-off among various existing data
structures.  It has an advantage in parallel construction and batch insertions,
scaling especially well to processing large batches, as well as outperforming
all other tree-based baselines on queries.  As a result, \ourtree demonstrates
consistent gains on real-world geometric and graph workloads
(\cref{sec:kv_store} and \cref{sec:temporal}). We release our anonymous code~\cite{pabcode}.

\section{Preliminaries} \label{sec:prelim}

This section reviews the analysis models, background on \btree{s}, and key
subroutines that we will use to prove theoretical guarantees on the \ourtree{}
operations.

In this paper, we use $n$ to denote the input size or the tree size. We write
$\log n$ for $\log_2(n+1)$, and $\log_B n$ for $\log_B(n+1)$.

\myparagraph{Computational Model.}  We assume a shared-memory multiprocessor that
supports fork-join parallelism with binary
forking~\cite{BL98,arora2001thread,blelloch2020optimal}.  A thread is a
sequential RAM 
with a fork instruction
that creates two parallel child threads and suspends the parent until both finish; 
a parallel for-loop compiles to a logarithmic number of forks.  We analyze our
algorithms using the work-span model, where the work is the total number of
operations in the algorithm, and the span is the longest dependence chain in the
parallel computation graph. On $p$ processors, a randomized work-stealing
scheduler executes a computation with work $W$ and span $S$ in $W/p+O(S)$ time
with high probability in $W$~\cite{BL98,arora2001thread,gu2022analysis}.

We use \defn{I/O cost} to refer to cache complexity, and analyze it using the
ideal-cache model~\cite{Frigo99,aggarwal1988input}: a cache of size $M$ above
an unbounded memory, exchanging size-$B$ blocks at unit cost under an optimal
offline policy. The cache complexity is the total number of block transfers. We
also use $B$ for the block size of our \ourtree, which can be any value in $\Theta(B)$.

%
%
%
%

\myparagraph{\btree{s}.}  The \bplustree{} is a canonical data structure for storing
ordered data.  Every node keeps a $\Theta(B)$ fan-out so that it aligns with a
hardware access unit (e.g., a cache line or a disk block).  All entries are
stored in the leaves, and all leaves are at the same depth.  Each leaf contains
a sequence of entries in sorted order.  An interior node contains a sequence of
child pointers and a sequence of splits that separate its children.  Every
interior node except the root has between $\lb$ and $\ub$ child nodes.
A leaf can hold more entries since it stores no child pointers.  However, for
simplicity we assume that a leaf also holds between $\lb$ and $\ub$
entries.

A \bplustree{} with $n$ entries has height $O(\log_B n)$. It can be constructed
sequentially with $O(n\log_B n)$ work and $O(n/B\cdot\log_B n)$ I/O cost.
During insertion, a node overflows if its size exceeds $\ub{}$.  This
overflow is resolved by splitting the node into two and promoting the midpoint
key to its parent.  Conversely, deleting an entry may decrease a node's size
below $\lb{}$. This triggers an underflow that is handled by merging the node
with an adjacent sibling and removing a key from the parent.  Both overflows and
underflows can cascade up the tree, requiring recursive rebalancing.

\section{Parallel Algorithms\label{sec:algorithms}}

\begin{figure}[t]
	\includegraphics[width=0.48\textwidth]{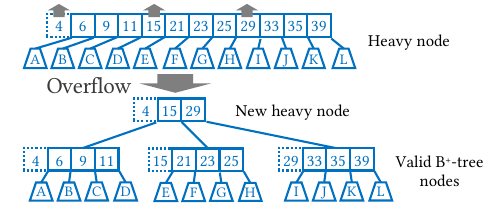}
	\caption{
		\textbf{Illustration of the \overflow{} primitive. }
		Assume $B=2$. Then \textit{\overflow} adjusts each node to a fan-out of $2B=4$. Trapezoids with uppercase letters represent subtrees.
		The virtual leading keys are marked in dotted boxes.  The resulting subtrees
		are all valid and at the same level.  }
	\label{fig:overflow}
\end{figure}

\subsection{\pbtree{} Structure and Primitives}

In this section, we present the design of \pbtree{}. A \pbtree{} is
conceptually a \bplustree{}, where all data entries reside in the leaf nodes
and internal nodes store \emph{\separator{s}} as routing keys.  Unlike
a standard \bplustree{}, where a node with $b$ subtrees contains $b-1$
\separator{s}, an internal node in the \pbtree{} always maintains $b$ \separator{s}
and the same number of subtrees, where each subtree follows a key.  A key and
the subtree following it are \defn{paired}.  Such a node can be viewed as
a regular \bplustree{} node augmented with a \emph{leading key}, which is the
smallest key in its subtree.  We always augment the entire dataset with
a leading key $-\infty$ to ensure correctness.
\cref{fig:overflow} and \cref{fig:insert_illustration} both illustrate this scheme.  The
leading key simplifies the algorithm design, since handling internal and leaf
nodes can then follow the same logic.  It also simplifies our proofs.  In
practice, there is no need to store the leading key explicitly, since it can
always be retrieved from the recursive calls.

A valid node has size in $[\lb,\ub)$.  Any $cB$ with $c>2$ works as the upper
bound and guarantees theoretical and practical efficiency.
We fix $c=4$ for simplicity, and later show in our proofs how it bounds the costs.


\myparagraph{\titlecap{\heavynode{}}.}  The concept of a \defn{\heavynode{}}
simplifies the description of the \pbtree{} algorithms.  A \heavynode{}
encapsulates the transient state of an invalidated \bplustree{} node.
We allow a heavy node to have any size of at least $\lb$, with no upper bound,
but all its subtrees must be valid \bplustree{s} of equal height.
This property is crucial, as structural invalidities are isolated and resolved
locally within the node, streamlining the overall correctness justification
(see \cref{fig:overflow} for an illustration).  Specifically, a \heavynode{}
$H$ maintains its $\size{}$, also denoted by $|H|$, and two arrays $\nds{}[]$
and $\seps[]$, each of size $|H|$.  Here, $H.\seps[i]$ is the smallest key in
$H.\nds{}[i]$.  A valid tree node is therefore a \heavynode{} with size in
$[\lb,\ub)$.  We use the \heavynode{} as the core operational unit in our
algorithms.

\begin{algorithm}[t]
	\fontsize{9pt}{9.5pt}\selectfont
	\caption{\pbtree{} Primitive: \overflow\label{algo:primitive}}
	\SetKwFor{parFor}{parallel-for}{do}{endfor}
	\SetKwFor{parForEach}{parallel-foreach}{do}{endfor}
	\SetKw{MIN}{min}
	\SetKw{MAX}{max}
	\SetKwProg{MyFunc}{Function}{}{end}
	\SetKwInOut{Note}{Note}
	\SetKwInOut{Parameter}{Parameter}
	\SetKwFor{inParallel}{in parallel:}{}{}
	\SetKwFor{hnode}{A \heavynode{} maintains}{}{}
	\SetKwSty{textbf}
	\SetArgSty{textnormal}
	\SetFuncSty{textnormal}

	\let\oldnl\nl
	\newcommand{\nlnonumber}{\renewcommand{\nl}{\let\nl\oldnl}}

	\DontPrintSemicolon
	\nlnonumber \hnode{}{
		\nlnonumber \boldmath $\size{}$\unboldmath: size of the heavy node, also denoted by $|\cdot|$\\
		\nlnonumber \boldmath $\seps{}$\unboldmath: an array of $b$ \separator{} keys\\
		\nlnonumber \boldmath $\nds{}$\unboldmath: an array of $b$ \pbtree{} subtrees\\
	}
	\codeskip
	\tcp{Resolves overflow in \heavynode{} $H$ and return a parent \heavynode{}.}
	\MyFunc{\upshape $H'\gets$ {\overflow{$(H)$}}}{
	\lIf{$H$ is a valid node}{return $H$}
	$b'\gets \max(\lfloor |H|/\midb \rfloor,1)$\tcp*[f]{to be split into $b'$ valid \pbtree{} nodes}\\
	$H'.\size{}\gets b'$\\
	\parFor(\tcp*[f]{partition $H$ into $b'$ groups}){$i\gets 0$ to $b'-1$} {
	\tcp{Group $i$ contains the $s$-th to the $e$-th \sep{} and \nd{} in $H$, each with similar sizes (differing by at most 1)}
	$s\gets \lfloor i\cdot b/b'\rfloor$, $e\gets \lfloor (i+1)\cdot b/b'\rfloor$\\
	Create a \pbtree{} node $v$ using subtrees $H.\nds[s..e)$ and \separator{s} $H.\seps[s..e)$\\
		\tcp{Set the $i$-th node and \separator{} in return heavy node $H'$}
	$H'.\nds[i]\gets v$\\
	$H'.\seps[i]\gets$ the first key in $v$\\
		}
		\Return $H'$ \tcp*[f]{Return the parent heavy node}
	}
\end{algorithm}

\myparagraph{The \titlecap{\overflow{}} Primitive.}  Our algorithm follows the
standard overflow and underflow rebalancing scheme of \bplustree{s}.  We first
introduce the \overflow{} primitive (\cref{algo:primitive}), which simulates the handling of
an overflow.
It restructures a \heavynode{} into a sequence of valid nodes at the same level, each of size about $\midb$, and generates a higher-level \heavynode{}.
In detail, the \overflow{} first check whether $H$ is a valid node, and if so, returns it directly.
This is consistency with the definition of a \heavynode{} since a valid node is also a \heavynode{}.
Otherwise, it partitions the $b$
subtrees and \separator{s} of a heavy node $H$ into $b'\mathit{=}\lfloor b/\midb\rfloor$ groups,
each of size $\lceil b/b'\rceil$ or $\lfloor b/b'\rfloor$. It then packs each
group into a standard node and extracts its leading key to form $b'$
\separator{s}.  Crucially, even when $H$ does not overflow, at least one
separator is still promoted: the \overflow{} packs $H$ into a single node and promotes the key paired with it, so the operation applies uniformly.
The resulting
$b'$ subtrees and \separator{s} form a smaller \heavynode{}, which is then passed to the parent
(\cref{fig:overflow}).

The \overflow{} primitive extends the sequential overflow process naturally, and
also admits parallelism: the grouping depends only on local
indices, so all subtrees and \separator{s} are processed in parallel,
resolving an overflow of any size in a single step.
With this simple primitive, our insertion algorithm
largely reduces to applying this primitive repeatedly, and our deletion
algorithm also implicitly leverages it.

Below, we detail the complete algorithms based on the \heavynode{} and the \overflow{} primitive.



\begin{figure*}[t]
    \vspace{-2.3em}
	\includegraphics[width=\textwidth]{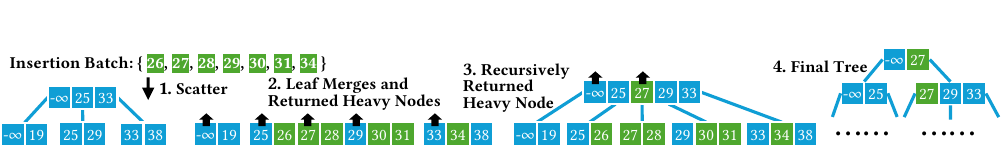}
	\caption{
	\textbf{An illustration of the batch insertion algorithm with $B=1$.}  The algorithm runs in an asynchronous divide-and-conquer approach, and the output of each subtree is always a heavy node.  The two heavy nodes each with a size over 4 apply the \overflow{} function.}
	\label{fig:insert_illustration}
    \vspace{-1em}
\end{figure*} 

\begin{algorithm}[t]
	\fontsize{9pt}{9.5pt}\selectfont
    \SetKwSty{textbf} 
    \SetArgSty{textnormal}
    \SetFuncSty{textnormal}
	\caption{Batch Insertion\label{algo:insert}}
	\SetKwFor{parForEach}{parallel-foreach}{do}{endfor}
	\SetKwFor{parFor}{parallel-for}{do}{endfor}
	\SetKwProg{MyFunc}{Function}{}{end}
	\SetKwInOut{Note}{Note}
	\SetKwInOut{Parameter}{Parameter}
	\SetKwFor{pardo}{In Parallel:}{}{}
	\KwIn{A sequence of entries $E$ and a \pbtree{} $T$.}
	\KwOut{A \pbtree{} with $E$ inserted.}
	\DontPrintSemicolon

        \vspace*{.5em} \tcp{Insert the entries $E$ into $T$ and
          maintain the \bt{} properties}
        \MyFunc{\upshape{\batchinsert{$(T,E)$}}}{
		Sort the entries in $E$ by their keys\label{line:insert-sort}\\
		$H\gets$\textsc{InsertSorted$(T,E)$}\tcp*[f]{Insert new entries into leaves}\\
		\While{$H.\size{}\ge \ub$}{
            $H\gets$ \overflow$(H)$
        }
        \Return $H$
              }

	\tcp{Insert sorted entries $E$ into $T$; return updated node, possibly heavy}
	\MyFunc{\upshape{\textsc{InsertSorted$(T,E)$}}}{
		\lIf{$E$ is empty}{
			\Return $T$ \tcp*[f]{Unchanged}\label{line:insert:empty}
		}
		\If{$T$ is a leaf}{
			Insert $E$ into $T$ in place\\
            \tcp{Use \overflow{} to split the tree when necessary, which also identifies the upper-level \separator{s} to return to the parent}
			\Return \overflow$(T)$ \label{line:insert:append_leaf}
		}
		\tcp{Otherwise, $T$ is an interior node; continue recursive insertion}
        $b\gets T.\size$\\
		$P[0..b)\gets$ Slices of $E$ partitioned by $T.\seps$ \label{line:insert:split} \\
        Initialize $\pufseq$ as a sequence of $b$ \heavynode{s}, which will hold the results returned by the $b$ recursive calls\\
		\parFor(\tcp*[f]{Recursively insert}){$i\gets 0$ to $b-1$}{
			$\pufseq[i]\gets$\textsc{InsertSorted$(T.\nds{}[i], P[i])$}
		}
        Initialize $H'$ as a heavy node, which will represent the state of $T$ after absorbing all insertions \\
        $H'.\nds\gets$ concatenate all $\pufseq[\cdot].\nds[]$ in order \\
        $H'.\seps\gets$ concatenate all $\pufseq[\cdot].\seps[]$ in order\\
        \tcp{Resolving possible overflow and pass new \separator{s} to parent}
        \Return \overflow$(H')$\label{line:insert:append_interior}
	}

\end{algorithm}

\subsection{Batch Insertion and Construction}\label{sec:insert}

We now present our parallel batch update algorithms for \pbtree{s}.  We consider
the batch-parallel setting, where a batch of entries is inserted into or deleted
from the current \pbtree{} $T$. We start with the insertion algorithm.

With the \overflow{} primitive, the insertion process is relatively
straightforward.  We present the algorithm in \cref{algo:insert} and illustrate it in \cref{fig:insert_illustration}. The algorithm inserts a batch of entries $E$ into a \pbtree{} $T$.  We first sort all keys in $E$, then follow a divide-and-conquer scheme and pass all entries to the root
by calling the helper function \textsc{InsertSorted}.  This function takes a
subtree~$T$ and inserts the sorted entries $E$ into it.
It returns a \emph{\heavynode{}} $H$ holding the subtrees replacing $T$, each paired with its leading key.
By definition, the returned node is a
sequence of $b'$ \separator{s}, each followed by a subtree.  Here $b'$ can be
any integer no smaller than $\lb$.

\textsc{InsertSorted} has two base cases:  1) an empty batch leaves the subtree
unchanged, and 2) a leaf merges the entries $E$ in place and calls \overflow{}.
Even though a leaf stores no subtrees, the \overflow{} logic applies identically, with the entries playing the role of subtrees; i.e., it splits the leaf into new leaves of about $\midb$ entries each and promotes the leading keys.
At an internal node, we partition $E$ by the \separator{s} in $T$ and
recurse in parallel on every pair of subtree and slice.  Each recursive call
returns a \heavynode{}.  Because every \heavynode{} retains its leading key,
these results concatenate directly into one transient \heavynode{} $H'$.
A final call to \overflow{$(H')$} further partitions $H'$ into valid subtrees and returns them to the parent as a \heavynode{} (line~\ref{line:insert:append_interior}).

\myparagraph{Parallel Construction.} We construct a \pbtree{} in parallel by inserting a batch of entries $E$ into an empty tree.
This is equivalent to viewing the entire batch as one large \heavynode{} and applying \overflow{} to it repeatedly, until a valid root is generated.

\myparagraph{Theoretical Analysis.}
We can now show the bounds for the \ourtree insertion algorithm.

\begin{theorem}
\label{thm:btree_insertion}
Given a \ourtree of size $n$, a batch insertion of $m$ entries takes
$O(Bm \log_B (m+n))$ work, $O(m \log_B(m+n))$ I/O cost, and
$O(\log n \log(m+B))$ span.
\end{theorem}

The I/O cost per entry is $O(\log_B(m+n))$, which aligns with the sequential I/O
cost of B/B$^{+}$-trees.  The work bound is larger by a multiplicative factor of
$O(B)$.
Given the high compute throughput and limited memory bandwidth of modern processors, our \pbtree{} is rarely bottlenecked by computation, unless the tree fits in cache.
The span is polylogarithmic.

The analysis of I/O cost is based on the following fact.

\begin{fact}\label{fact:tree-node}
Given a \ourtree with $n$ leaves and height no more than $\lceil\log_B n\rceil$, the total number of nodes among any $m$ leaves and their ancestors is $O\left(m \log_B \frac{n}{m}\right)$.
\end{fact}
Because the fan-out of a \ourtree is at least $B$, this height is an
overestimation.
We pessimistically assume that all nodes in the top $\lceil\log_B m\rceil$ levels are ancestors of the $m$ leaves, which gives $O(m)$ nodes in total. Each of the remaining $\lceil\log_B n\rceil - \lceil\log_B m\rceil = O(\log_B(n/m))$ levels contains at most one ancestor per leaf, hence at most $m$ nodes, which leads to the stated bound.

We can now prove the batch insertion bounds in \cref{thm:btree_insertion}.

\medskip
\begin{proof}[Proof of \cref{thm:btree_insertion}]
  We can directly obtain the I/O bound from Fact~\ref{fact:tree-node}.  Since each
  node in a \ourtree aligns with $O(1)$ memory blocks (cache lines), inserting
  $m$ entries into a tree of size $n$ can touch at most the number of nodes
  bounded by Fact~\ref{fact:tree-node}.  This corresponds to
  $O(m \log_B ((m+n)/m))$ cache line reads and writes.  Note that this bounds
  the nodes touched in the output tree.
  The input tree (before
  modification) must also be accounted for, but in an insertion operation, the
  number of tree nodes in the input tree is less than or equal to that
  of the output tree.
  Adding both together yields the stated
  $O(m \log_B ((m+n)/m))$ I/O bound.  We also need to sort the input
  entries (line~\ref{line:insert-sort}) with $O(m\log_M m)$
  I/Os~\cite{blelloch2010low}, and adding all pieces gives the overall
  $O(m \log_B(m+n))$ I/O bound.

  \hide{ The work bound consists of two parts: finding the insertion locations
    for the entries, and modifying the tree structure.  The first part contains
    searching for $O(\log_B(n+m))$ levels, each level including a $O(\log B)$
    binary search.  The total work of identifying the exact insertion location
    in an leaf is $O(\log(n+m))$.  For the second part, inserting the entries to
    the leaves takes $O(B)$ work per element.  For the overflow, as in the
    sequential counterpart, any tree node get split } The work bound is obtained
  by multiplying the I/O cost by $B$.  As each node is accessed and processed,
  operations such as scanning, merging, and concatenating all take work linear
  in the input size, which is $O(B)$.  Therefore, the total work excluding the
  sorting part is $O(Bm \log_B (m+n))$.  Sorting takes
  $O(m\log m)=O(m\log (m+n))=O(Bm \log_B (m+n))$; thus, the total work remains
  $O(Bm \log_B (m+n))$.

  Finally, we analyze the span bound. The insertion algorithm recurses for
  $\log_B n$ levels.  At each level, the parallel-for loop of size $B$ produces
  a span of $\log B$.  Multiplying these two factors yields a span of $\log n$.
  At each level, the work for operations including merge and
  concatenate can be trivially parallelized with a span logarithmic in the
  input size.  Since the worst-case input size is $m+B$, the span per
  level is bounded by $O(\log(m+B))$.  Together, the total span is bounded by
  $O(\log n \log(m+B))$, leading to the stated bound.
\end{proof}

\setlength{\algomargin}{0em}
\begin{algorithm}[t]
	\fontsize{9pt}{9.5pt}\selectfont
	\caption{Batch Deletion\label{batchdelete}}
	\SetKwSty{textbf} 
    \SetArgSty{textnormal}
    \SetFuncSty{textnormal}
    \SetKwFor{parForEach}{parallel-foreach}{do}{endfor}
	\SetKw{MIN}{min}
	\SetKw{MAX}{max}
	\SetKwProg{MyFunc}{Function}{}{end}
	\SetKwInOut{Note}{Note}
	\SetKwFor{inParallel}{in parallel:}{}{}
	\SetKwFor{parFor}{parallel-for}{do}{endfor}
	\SetKwInOut{Parameter}{Parameter}
	\KwIn{A sequence of unsorted entries $E$ and a \pbtree{} $T$.}
	\KwOut{A \ourlib{} with $E$ deleted.}
	\DontPrintSemicolon \vspace{.5em}
        \MyFunc{\upshape{\batchdelete$(T, E)$}}{
		Sort the entries in $E$ by their keys\\
		\textsc{DeleteSorted$(T,E)$}\tcp*[f]{Remove entries from $T$}\\
		Group the dropped subtrees in line \ref{line:delete:return-invalid} by height, and reinsert each group into $T$ using \cref{algo:insert}\label{line:delete:reinsert}\\
              }

        \vspace{1em}
        \MyFunc{\upshape{\textsc{DeleteSorted}$(T, E)$}}{
                \lIf{$E$ is empty}{
                        \Return $T$
                }
                \lIf{$T$ is a leaf}{
                        Remove $E$ from $T$, and \Return $T$
                }
                \tcp{$T$ is an interior node; continue the recursive deletion.}
                $P\gets$ Slices of $E$ partitioned by $T.\seps{}$\\

        $b\gets T.\size$\\
        Initialize $\nseq$ as a sequence of $b$ \pbtree{} nodes, which will hold the results returned by the $b$ recursive calls\\
		\parFor(\tcp*[f]{Recursively delete}){$i\gets 0$ \KwTo $b-1$}{
			$\nseq[i]\gets$\textsc{DeleteSorted$(T.\nds{}[i], P[i])$}
		}
        \If{\upshape the number of non-empty $\nseq[i]$ is less than $B$\label{line:delete:check-valid}}{Drop non-empty $\nseq[i]$ to be reinserted later, and \Return{$\emptyset$}\label{line:delete:return-invalid}}
        Create a \pbtree{} node $N$ by concatenating $\nseq$, and \Return{$N$}\label{line:delete:return-valid}
}
\end{algorithm}

\subsection{Batch Deletion}
Classic underflow logic is inherently sequential.
An underflowing node is repaired by borrowing from or merging with a sibling, so the outcome depends on that sibling rather than on the node alone.
A batch of deletions may trigger many underflows at once, and their repairs would conflict: the sibling borrowed from may itself underflow, or may be deleted entirely.
Resolving these interactions requires serializing the dependencies among the repairs, which blocks parallelism.

Our deletion algorithm (\cref{batchdelete}) avoids these interactions entirely, and the design is straightforward.
It follows the same divide-and-conquer approach as the insertion algorithm to maximize efficiency.
The algorithm first checks whether it has reached a base case: an empty entry list or a leaf node.
Otherwise, it recursively deletes the entries in the subtrees and collects the updated subtrees.
If at least $B$ subtrees survive (line~\ref{line:delete:check-valid}), we build a node wraps them and return it (line~\ref{line:delete:return-valid}).
If fewer survive, node $T$ underflows. The algorithm removes it and returns $\emptyset$ to mark its absence (line~\ref{line:delete:return-invalid}).
The surviving subtrees of the underflow node are set aside and later reinserted into the global \pbtree{}.
This involves collecting all dropped subtrees into an array in parallel, using a prefix sum over per-node counts to compute offsets, and then group subtrees by height for reinsertion.


Reinserting these subtrees uses the same insertion algorithm (\cref{algo:insert}).
The only difference is that the inserted items are subtrees instead of leaf entries.
The logic of \cref{algo:insert} still applies to both cases, 
because inserting a height-$h$ subtree into a node at height $h+1$ is the same operation as inserting an entry into a leaf.

\myparagraph{Theoretical Analysis.} 
We now analyze the complexity of batch deletion, which is summarized in \cref{thm:batch_delete}.
The high-level idea is to amortize the cost of reinsertion over the deletions that caused it.
A node is created with at least $2B$ children and is dropped only when fewer than $B$ elements remain, so at least $B$ deletions must happen below it before triggering a reinsertion. Those $B$ deletions together have paid $O(B\log_Bn)$, which covers reinserting the at most $B$ surviving subtrees.

\begin{theorem}
\label{thm:batch_delete}
Given a \pbtree{} of size $n$, batch deleting $m$ entries takes $O(Bm \log_B n)$ amortized work, $O(m \log_B n)$ amortized I/O cost, and $O(\log^2 n \log (m+B))$ span.
\end{theorem}

\begin{proof}
We analyze the complexity by dividing the deletion process into two phases: the \textsc{DeleteSorted} execution and the subsequent subtree reinsertions.

We first bound the I/O cost.
The \textsc{DeleteSorted} phase recursively routes the deleted entries down to the tree leaves and removes them.
Since this phase requires no structural repairs and follows a simpler access pattern than \textsc{InsertSorted}, its I/O cost is $O(m \log_B n)$.

The reinsertion phase requires an amortized analysis.
By the definition of the \pbtree{} structure, any valid tree node is initially constructed with a fan-out of at least $2B$.
During deletion, a node is dropped only if the number of its remaining subtrees falls below the degree bound $B$.
This implies that at least $B$ subtrees must have been previously dropped to trigger the drop of the current node.
Generalizing this property, we know that for a subtree at height $h$ (where leaves are at $h=0$) to underflow and be dropped, at least $B^{h}$ entries must have been deleted from it.

When such a subtree drops, we collect its remaining valid subtrees and reinsert them into the \pbtree{}.
Since there are at most $B$ subtrees left, reinserting them incurs an I/O cost of $O(B \log_B n)$.
We can charge this cost to the previously deleted entries in this subtree following a standard amortized analysis:
each previously deleted entry pays $O((B \log_B n) / B^h) = O((\log_B n) / B^{h-1})$ for the restructuring of its height-$h$ ancestor.
Summing this geometric series across all heights from $h=1$ to $O(\log_B n)$ yields an amortized cost of $O(\log_B n)$ per entry.
Combining this with the $O(\log_B n)$ per-entry I/O cost from the \textsc{DeleteSorted} phase, we obtain an overall I/O cost of $O(m \log_B n)$ for batch deletion.

The work bound analysis is similarly structured.
In the \textsc{DeleteSorted} phase, the algorithm performs simple data scanning and key routing without structural repairs, and its work is bounded by a factor of $B$ times its I/O cost.
For the reinsertion phase, because the algorithm directly reuses the batch insertion subroutine, its work likewise remains a factor of $B$ times its I/O cost.
Scaling the overall amortized I/O cost by $B$ yields an amortized work complexity of $O(Bm \log_B n)$.

Finally, we analyze the span bound.  The span of the \textsc{DeleteSorted} phase
is bounded by that of the \textsc{InsertSorted} phase.  However, rather than
inserting all dropped subtrees simultaneously, we group the dropped subtrees by
height and iteratively reinsert each group level by level.  This iterative
approach adds an extra $O(\log_B n)$ factor to the span depth.  Multiplying the
span of the insertion algorithm by this factor yields an overall span bound of
$O(\log^2 n \log (m+B))$.  This extra log factor could be removed by reinserting
all heights together, at the cost of redesigning insertion. Since the span is
already polylogarithmic and parallelism is strong in practice, we keep the
simpler design.
\end{proof}

\subsection{Optimizations}
To better accommodate dynamic workloads, \pbtree{} incorporates a probabilistic leaf-sorting strategy controlled by the \textit{leaf randomness ratio} $\alpha$: an update leaves a leaf unsorted with probability $\alpha$, and re-sorts it otherwise. Accordingly, queries scan an unsorted leaf linearly instead of using a binary search. For write-heavy workloads, this reduces the work of batch updates (e.g., insertions can directly append), at the cost of slightly slower queries. Note that the I/O cost is not affected because once a leaf node is loaded into cache, scanning it incurs no additional cache misses. We use $\alpha=0.5$ by default, but a user can adjust it according to the update/query ratio of their target workload. We experimentally evaluate this strategy in~\cref{sec:exp:randomness}.



%
\begin{figure*}[t]
	\includegraphics[width=1.0\textwidth]{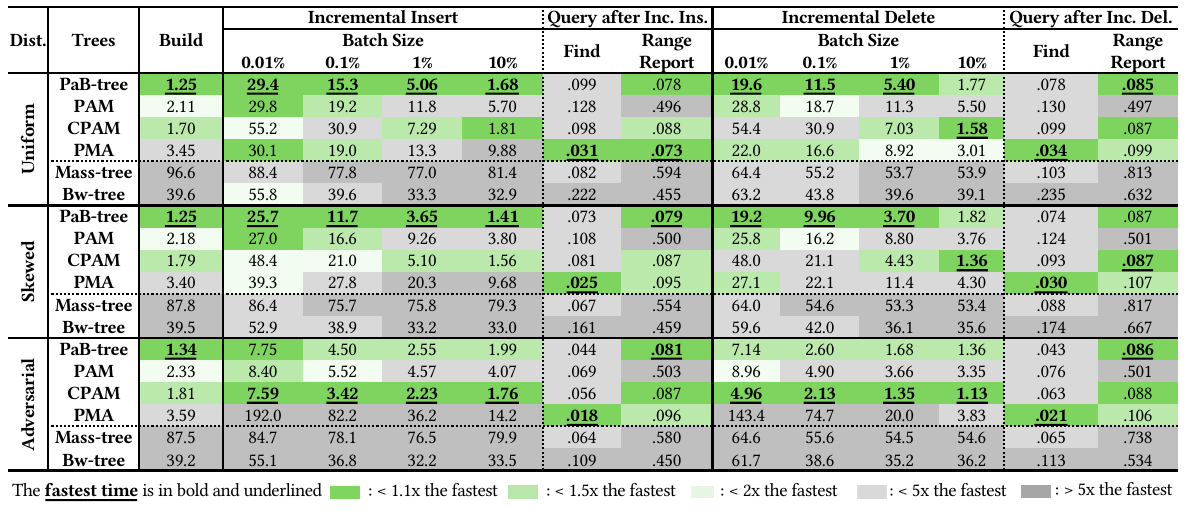}
	\caption{\textbf{Running time (in seconds) on synthetic data. Lower is better.}
		The fastest running time for each dataset is bold and underlined.
		We use different colors to mark results within 1.1$\times$, 1.5$\times$, 2$\times$, 5$\times$ and $>5\times$ of the fastest time.
		Detailed settings for build, queries, and incremental insertion/deletion are introduced at the beginning of \cref{sec:synthetic}.
	}
	\vspace*{-.6em}
	\label{fig:summary}
\end{figure*}


\section{Experiments\label{sec:exp}}

We conduct comprehensive experiments to evaluate the performance of \ourtree{}
against several baselines on both synthetic and real-world datasets.  Our
results show that \ourtree{} offers a clear advantage in batch updates while
delivering competitive or better query performance compared to existing
tree-based baselines.

\myparagraph{Experimental Setup.}  We perform experiments on a machine with
112 cores (224 hyper-threads), four Intel Xeon Platinum 8176 CPUs, and
1.47 TB RAM.  We implement \ourtree{} in \texttt{C++} and compile it with GCC
15.2.0 with the \texttt{-O3} optimization flag.  We use ParlayLib~\cite{blelloch2020parlaylib} for
fork-join parallelism.  The reported numbers are the average of 3 runs after a
warm-up run.  We use block size $B=64$\footnote{Evaluation for different block sizes is shown in \ifconference{the supplemental material.}\iffullversion{\cref{app:block-size}.}} and set the leaf randomness ratio
$\alpha= 0.5$.  The entries are sorted when the tree is initialized.  We release
our code at~\cite{pabcode}.

\myparagraph{Baselines.}  We compare with several state-of-the-art
batch-parallel and concurrent data structures.
For the parallel case, we evaluate (1) \defn{\pam{}}~\cite{sun2018pam}, a parallel binary search tree, (2)
\defn{\cpam{}}~\cite{dhulipala2022pac}, which improves the cache-friendliness of \pam{} with
leaf wrapping, and (3)
\defn{\pma}~\cite{WheatmanXu18,wheatman2023optimizing}, a Packed Memory Array that stores all entries
in a single contiguous sorted array, optimized for queries.
For the concurrent case,
we evaluate (1) \defn{\masstree{}}~\cite{mao2012cache} and (2) \defn{\bwtree{}}~\cite{wang2018building},
two well-known in-memory concurrent \btree{} implementations.

\subsection{Evaluation on Synthetic Benchmark\label{sec:synthetic}}

\myparagraph{Setup.}  We test different batch update patterns and key
distributions.  We test both the static and dynamic cases.  In the static case,
we measure the time to construct a tree with $10^9$ entries.  In the dynamic
case, we measure both the incremental insertion and deletion.  For insertion, we
divide $n=10^9$ entries into batches of size $m$ and insert them into the
tree $n/m$ times, starting from an empty tree.  For deletion, we first construct
a tree with all entries, then similarly remove them in batches until the tree is
empty.  We report the total time of all batches.  Under this workload, we also
measure the query performance when half of the batches are applied.
We evaluate membership (\find{}) and range queries
(\rreport{}): $10^8$ random \find{} queries and $5\times 10^4$ \rreport{} queries, with range
sizes spanning $10^2$ to $10^4$. The query implementations all directly follow
the standard \bplustree{} algorithms. Each query runs serially, with different
queries in parallel.
We also test the update time for a single batch, compare it with that of sequential point updates, and show the results \iffullversion{in \cref{app:single-batch,app:small-batch}}\ifconference{in the supplemental material}.

We use three key distributions for batches: uniform (\uniform{}), skewed
(\ske{}), and adversarial (\adv{}).  \uniform{} generates keys uniformly at
random within the range.  \ske{} generates keys from a normal distribution with
mean $0.5$ and standard deviation $0.1$, which simulates the real-world scenario
where the data in a batch has certain locality.  \adv{} sorts the keys before
dividing them into batches, so that each batch contains a consecutive range of keys.
This simulates the worst-case scenario for \ourtree{}, since the batch updates
always trigger the largest possible subtrees to overflow or underflow,
maximizing the cost of node splitting and reinsertion.  However, this is the
best case for \pam{} and \cpam{} since their algorithms reduce to simply merging
in or splitting out a subtree on the right spine.  We use 32-bit integer keys
and values, and summarize the results in \cref{fig:summary}.

\myparagraph{Construction.}  For tree construction, our \ourlib{} is
consistently faster than all baselines across all workloads.  This advantage
mainly comes from the shallower tree height and the blocked node layout of the
\btree{}, which enable fewer memory allocations and yield better cache behavior.
Both \pam{} and \cpam{} are binary trees with higher tree heights, and are
therefore up to 1.7$\times$ and 1.4$\times$ slower than \ourlib{}, respectively.
The \pma{} is up to 2.8$\times$ slower than \ourlib{}, possibly due to the
additional work required to scatter entries across the underlying array.  More
detailed hardware profiling is provided in \cref{sec:hardware}.  Both concurrent
indexes, \masstree{} and \bwtree{}, are orders of magnitude slower than
\ourlib{}, as concurrent insertion on a large batch incurs high node contention.

\myparagraph{Batch Updates.} For batch updates, \pbtree{} retains its advantage
in most cases.  It is always faster than the state-of-the-art BST \pam{}
(by up to 3.4$\times$).
In contrast, while \cpam{} also
leverages larger node sizes to improve I/O performance, it can be up to 1.9$\times$
slower than the standard BST in \pam{}, particularly on small batch sizes.
Altogether, on the \uniform{} and \ske{} cases, \pbtree{} is also almost always
faster than \cpam{}, by up to 2.8$\times$.

On large deletion batches (10\%) of \uniform{} and \ske{} data, and more markedly on \adv{} data, \cpam{}
outperforms \pbtree{}.  As noted earlier, the \adv{} workload consistently
inserts $m$ keys strictly larger than the tree's maximum value or deletes the
$m$ smallest keys.  This represents the worst-case scenario for \pbtree{}
because it is more prone to triggering node overflows and underflows; hence, we
include this workload as a stress test for \ourlib{}.  Interestingly, this is
the best-case scenario for \pam{} and \cpam{}, as their Join-based algorithms
simply merge or split trees along the rightmost path, reducing the cost to
$O(\log n+m)$, rather than $O(m\log n)$ in the general case.  In this
adversarial case, \pbtree{} is expectedly slower than \cpam{}, but only by
1.02--1.32$\times$ for insertions and 1.20--1.44$\times$ for deletions.  It also remains
faster than \pam{} at every batch size, by up to 2.0$\times$ for insertions and
2.5$\times$ for deletions.

Compared to the other baselines, including \pma{}, \masstree{}, and \bwtree{}, \pbtree{} is
faster in every batch-update case.  As an array-based
structure, \pma{} optimizes queries at the cost of update efficiency.  As a
result, \pbtree{} is on average 4.0$\times$ faster (up to 28.7$\times$) than
\pma{} on updates. Furthermore, both concurrent data structures perform worse
than batch-parallel ones on batch updates, since larger batches increase
contention rather than parallelism; \pbtree{} achieves an average speedup of
13.1$\times$ (up to 56$\times$) over \masstree{} and 7.9$\times$ (up to
27$\times$) over \bwtree{}.


\myparagraph{Queries.}  As expected, B-tree-like structures, including
\pbtree{}, \masstree{}, and even \cpam{}, perform better
than the strict BST \pam{} on all queries.  In \find{} queries, \pma{} has a
significant advantage over all other implementations, due to its array-based
layout and query-focused optimizations. However, this comes at the cost of
updates that are up to 28.7$\times$ slower than those of \pbtree{}.
Apart from \pma{}, \pbtree{} is almost always the fastest, or within 10\% of the
fastest, among the tree-based structures in \find{} queries.  For \rreport{}
queries, \ourtree{} is the fastest in four of the six cases and within 7\% of the fastest in the other two.  It achieves
speedups of up to 6.4$\times$ over \pam{}, up to 1.1$\times$ over \cpam{}, up to 1.2$\times$ over
\pma{}, and 5.6--9.6$\times$ over \masstree{} and \bwtree{}.

\subsection{Processing Real-world Geometric Data\label{sec:kv_store}}
\begin{figure}[t]
	\includegraphics[width=0.48\textwidth]{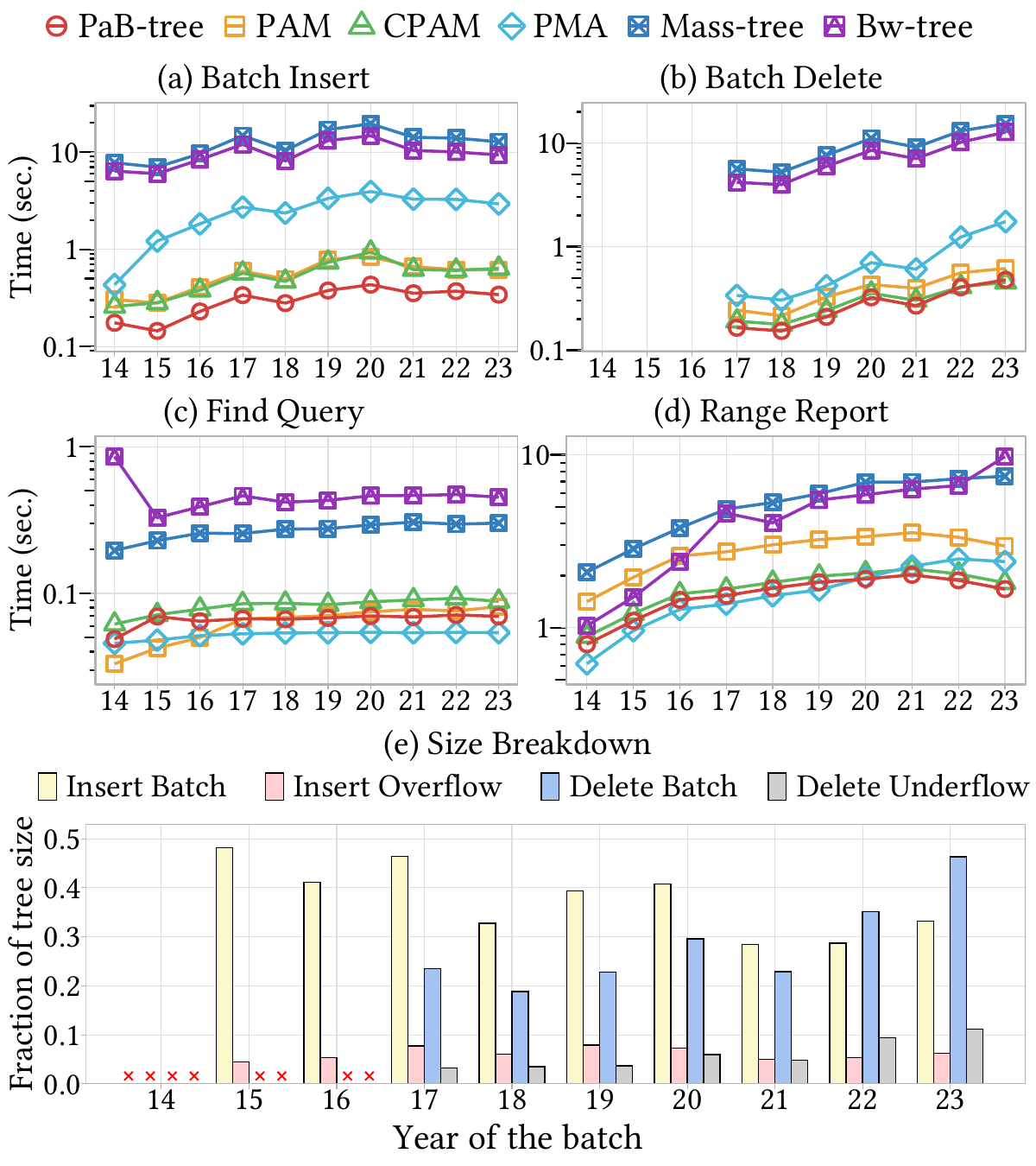}
	\caption{
		\textbf{Running time (in seconds) of batch updates and queries on the OSM dataset. Lower is better.}
		The inputs are batched by years from 2014 to 2023. In each year, we insert the batch in that year and then delete the batch from three years ago if applicable.
		After batch updates, we perform the queries on the current index.
		The \find{} query contains $10^8$ candidates, and the \rreport{} query contains $5\times 10^5$ queries with output size between $10^2-10^4$. Both queries are generated randomly.
		The yellow and blue bars depict the fraction of input batch sizes w.r.t the tree size in each year.
		The pink and grey bars show the fraction of overflow and underflow size w.r.t the tree size during the batch insertion and deletion, respectively.
              }
              \label{fig:osm}
\end{figure}

 We evaluate \ourlib{} as a general key-value store on a
real-world geometric dataset.

\myparagraph{Setup.}  We use \texttt{OpenStreetMap}
(\osm{})~\cite{haklay2008openstreetmap} as the benchmark.  For each record in
\osm{}, we use a 64-bit Morton code that encodes its geographical information as
the key, and a 32-bit integer for its timestamp as the value.  We divide the
1.35B inputs into 10 batches by their timestamps, from year 2014 to 2023.  We
simulate a sliding-window setting following the methodology
of~\cite{men2025parallel} with window length 3.
In each year, we insert the corresponding batch and, where applicable, delete
the batch from three years prior.  After the update for each year is applied, we
run membership queries for $10^8$ randomly picked entries and $5\times 10^5$ range-report
queries with output sizes in $[10^2, 10^4]$.

\myparagraph{Results.}  We summarize the results in \cref{fig:osm}.  \ourlib{}
is consistently faster than all baselines for batch updates.  For insertions,
\pam{} suffers from its large tree height, and \cpam{} must maintain a total
order within its leaves; these make them slower than \ourlib{} by 2.09$\times$
and 1.95$\times$, respectively.  For deletions, the gap between \pma{} and the
other parallel baselines is smaller, since deletion involves fewer
memory-allocation issues. Even though reinserted entries are up to $11.2$\%
of the total tree size, \ourtree 
stays up to $1.16\times$ faster than \cpam{}, showing
the practical value of its I/O efficiency.

For \find{} queries, \pma{} benefits from the absence of pointer chasing and is
the fastest among all baselines, 1.07--1.45$\times$ faster than \ourlib{}.  Its
advantage on \rreport{} queries is smaller, since the output size dominates the
query cost.  \pam{} provides better \find{} performance than \ourlib{} and
\cpam{} at small tree sizes, since the latter two must scan all entries in a
leaf to find the target key.  Meanwhile, without leaf blocking, \pam{} has poor
cache locality when reporting many leaf entries, which also makes it much slower
than the others for \rreport{} queries.  Both concurrent baselines, \masstree{}
and \bwtree{}, are orders of magnitude slower than \ourlib{} for both queries,
as locks block the queries.

\subsection{Processing Real-world Graph Data\label{sec:temporal}}
\begin{figure}[t]
	\includegraphics[width=0.48\textwidth]{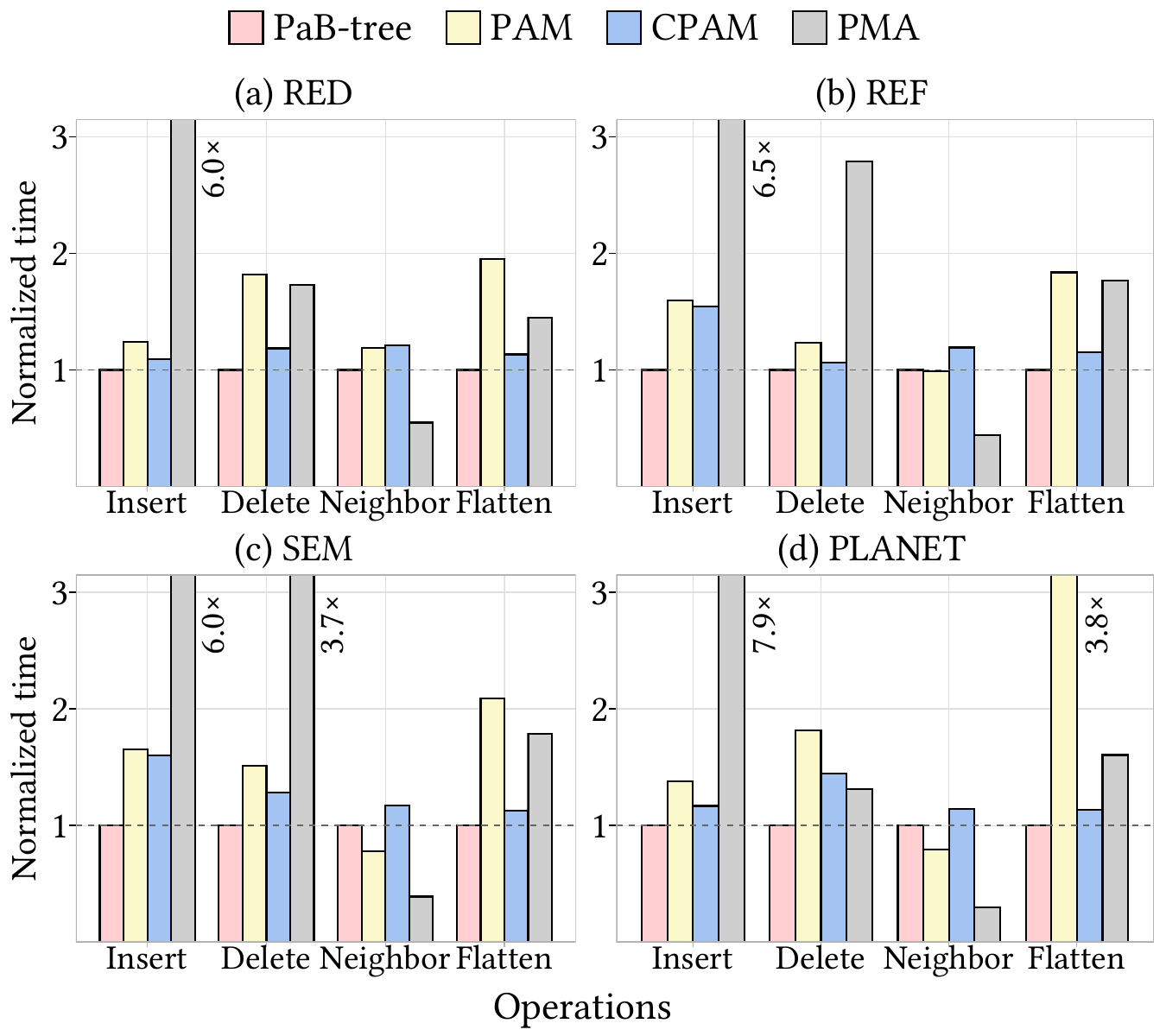}
	\caption{
		\textbf{Normalized running time of batch updates and queries on temporal graph datasets. Lower is better.}
		Time exceeds 3 times slower than \pbtree{} is chopped.
		Detailed settings can be found in \cref{sec:temporal}.
		For all operations, we measure the sum of running time across all batches and normalized to the \pbtree{}.
		The \neighbor{} query lists all neighbors for $10^7$ vertices with the largest degree.
		The \flatten{} query outputs all entries in a sorted order.
		\looseness=-1
	}
	\label{fig:temporal}
\end{figure}
 We evaluate \ourlib{} as a graph container on temporal
graphs.  Every edge in a temporal graph has a timestamp, and we group edges into
batches by their timestamps (in years) to simulate the evolution of the graph
over time.

\myparagraph{Setup.}  We compare the baselines on four benchmarks from
GraphTide~\cite{wheatman2026use}: \redd{} (746M edges, in 17 batches),
\reference{} (1.14B edges, in 27 batches), \sem{} (2.49B edges, in 229 batches),
and \planet{} (1.01B edges, in 20 batches).  The batch sizes generally grow over
the years.  For graphs without explicit insert or delete tags on their edges,
such as \redd{}, \reference{}, and \sem{}, we simulate the sliding-window
setting as in \secref{kv_store} with window length 3; otherwise, we form the
insertion and deletion batches according to those tags.  For each year, we first
apply the batched insertions, then the deletions.  After the batch updates in
each year, we run two fundamental primitives in graph algorithms: (1)
\texttt{Neighbor}$(V)$, which retrieves the neighbors of every vertex in $V$,
and (2) \texttt{Flatten}, which returns all edges in sorted order, forming a
compressed-sparse-row-like (CSR-like) array.  \texttt{Flatten} is widely used in
dynamic graph systems to extract the graph prior to query execution (e.g., the
\emph{flat snapshot} operation in Aspen~\cite{dhulipala2019low}).  We encode
each edge as a 64-bit integer and use it as the key, with another 32-bit integer
as the value for its timestamp.  We sum each operation's running time over all
years and normalize the total to \ourlib{}.

\myparagraph{Results.}  We summarize the results in \cref{fig:temporal}.  For
both batch insertion and batch deletion, \ourlib{} is the fastest among all
baselines, achieving 1.09--1.60$\times$ speedup over the fastest baseline
\cpam{} on insertion and 1.06--1.44$\times$ speedup on deletion.  This is
consistent with the findings in \cref{sec:kv_store}: we believe that keeping
leaf entries unsorted is the main reason for \ourlib{}'s strong batch-update
performance, while the overhead of reinsertion-based deletion remains modest in
practice.  As a static index, \pma{} is much slower than the others under highly
dynamic updates, due to frequent memory allocation and reconstruction; this
aligns with the findings on the synthetic datasets (\secref{synthetic}).

For graphs with a small average vertex degree ($\leq 10$), such as \planet{} and
\sem{}, \neighbor{} degenerates to \find{}, which favors \pam{}. For
large-degree graphs (\redd{}), this advantage diminishes. For \flatten{}, \pam{}
is the slowest due to excessive pointer chasing, while \pma{} must skip empty
slots in its underlying array; \ourlib{} and \cpam{} are comparable.

\subsection{Hardware Profiling\label{sec:hardware}}

\begin{figure}[t]
	\includegraphics[width=0.48\textwidth]{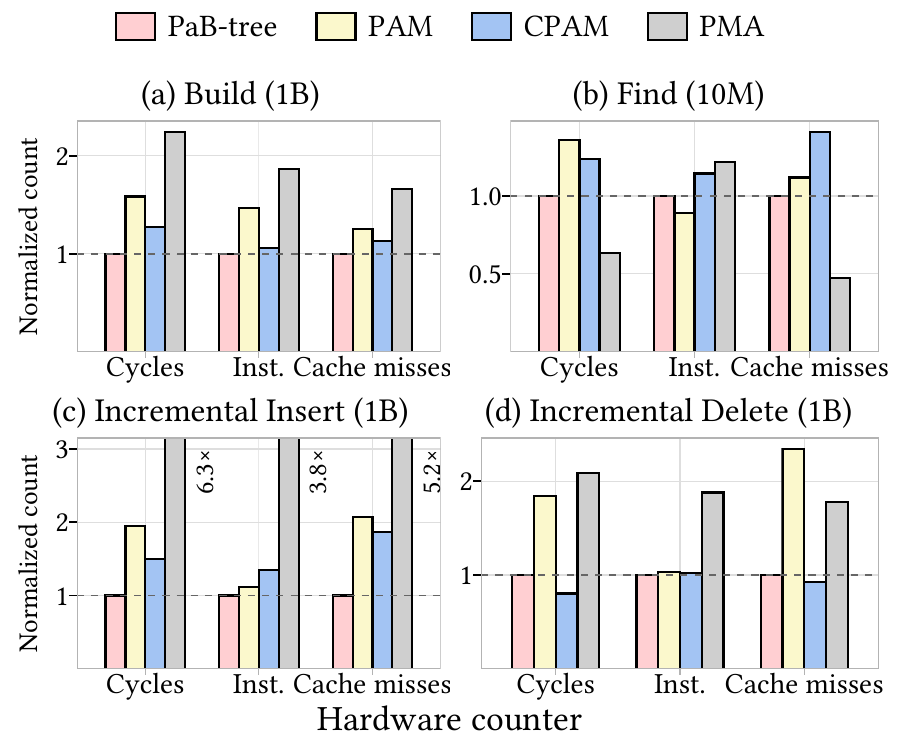}
	\caption{
		\textbf{Profiling the cycles, instructions and cache-misses of tree construction, incremental batch updates, and membership queries on \ske{} distribution. Lower is better.}
		The numbers are normalized to \ourlib{}.
		We use the \ske dataset with batch size of $10^7$, in total $10^9$ entries are processed.
	}
	\label{fig:cache}
\end{figure}
 We profile the CPU cycles, instruction
count, and cache misses of tree construction, incremental batch updates, and
\find{} queries on the \ske{} distribution, with $10^9$ entries and batches with size
of $10^7$.  The instruction count reflects the total work of the algorithm, and
the cache misses reflect its I/O efficiency.  We summarize the results in
\cref{fig:cache} and provide details in \ifconference{the supplemental material.}\iffullversion{\cref{app:hardware}.}

\ourlib{} benefits the most from its I/O efficiency, which gives it the best
performance on tree construction and batch insertion.  The low instruction count
from the algorithm's simplicity also contributes to the performance gains of
\ourlib{}.  Compared with tree indexes, the \pma{} incurs the fewest cache
misses for \find{} queries due to the absence of pointer chasing in inner tree
structures, which makes it the fastest for \find{} queries.  However,
incremental batch insertion frequently triggers memory allocation and
reconstruction of overflowed blocks; as a result, the large number of cache misses and
instruction counts make the \pma{} the slowest for batch insertion.  The leaf
blocking in \cpam{} enables it to have a lower tree height than the \pam{} and
therefore fewer cache misses needed to traverse the tree, making it consistently
faster than \pam{} for all operations.

\subsection{Evaluation of Randomness in Leaves\label{sec:exp:randomness}}
\begin{figure}[t]
	\includegraphics[width=0.48\textwidth]{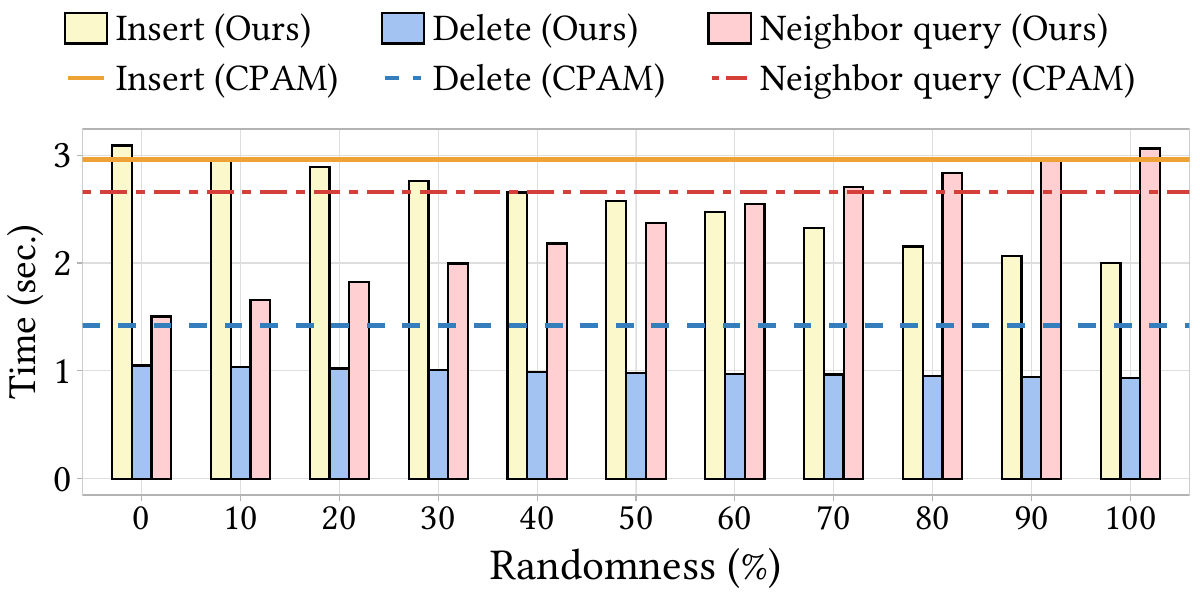}
	\caption{
		\textbf{Running time (in seconds) of batch updates and \neighbor{} queries w.r.t different randomness ratio in \ourlib{}. Lower is better.}
		We use the temporal graph datasets \planet{}. 
		The update edges are divided into 20 batches by their timestamps.
		For each timestamp, after the batch insertion and deletion, we run $10^7$ \neighbor{} queries.
		The bar shows the sum of the running time of the batch updates and the \neighbor{} queries.
		The corresponding times for \cpam{} are marked in lines.
	}
	\label{fig:randomness}
\end{figure}
 We evaluate the impact of sortedness in the leaves on
the performance of \ourlib{} on the temporal graph dataset \planet{} and
summarize the results in \cref{fig:randomness}.  The randomness ratio $\alpha$
controls the degree of randomness in the leaf entries, with $\alpha=0$
corresponding to a fully sorted leaf and $\alpha=1$ corresponding to a fully
random leaf. Sorted leaves improve query performance at the cost of more
expensive batch updates.

Both batch insertion and deletion become faster if we tolerate more randomness
in the leaf. Insertion is more sensitive to the ratio than deletion.  This
comes from the asymmetry between the two operations. With a larger randomness $\alpha$, an
insertion will likely just append new entries to an unsorted leaf, whereas
a sorted leaf requires merging, resulting in a more significant difference
between the two cases. On the other hand, a deletion has to filter the leaf in
one pass regardless of whether it is sorted, so $\alpha$ minimally affects its
leaf-level cost. The remaining effect of $\alpha$ on deletion comes from the
reinsertions it triggers, which follow the insertion path. Since the reinserted
subtrees are usually much smaller than the batch itself, $\alpha$ affects the total
deletion time much less than the insertion time.

A \neighbor{} query needs to locate the start and end of the range in the leaf. A
sorted leaf enables binary search, which is much faster than linear search in a
random leaf.  As a result, the query time increases as the randomness ratio
grows.  Overall, the randomness ratio $\alpha=0.5$ provides a good balance
between batch update and query performance.

\subsection{Scalability}
\begin{figure}[t]
	\includegraphics[width=0.48\textwidth]{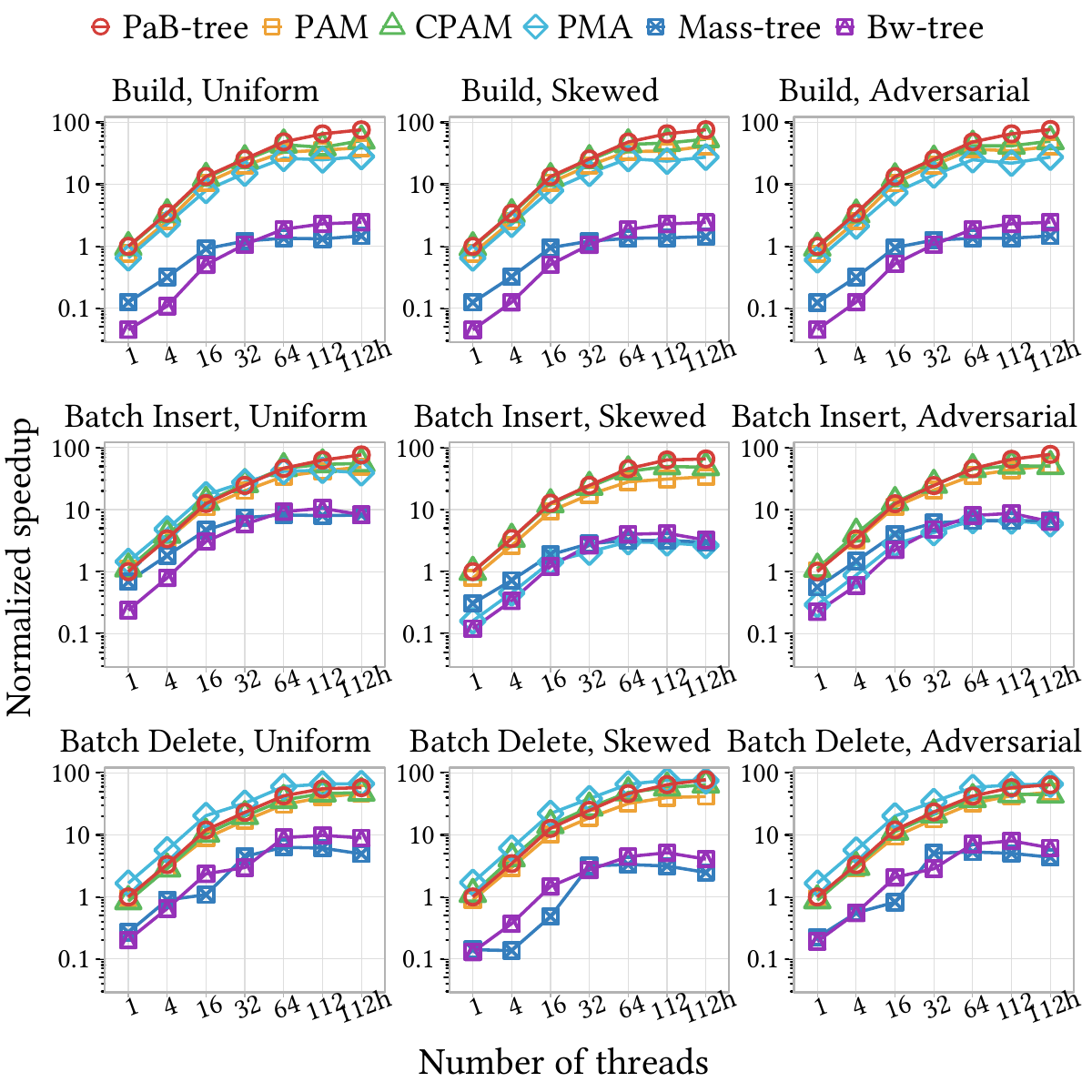}
        \caption{\textbf{Normalized parallel speedup of tree construction, batch
            insertion and batch deletion on different number of
            processors. Higher is better.}  The speedup is normalized to the
          running time on the \ourlib{} on 1 thread respectively.  The $10^9$
          entries, and batch updates apply a single batch of $10^7$ entries.
          ``112h'' means using all 112 cores (224 hyper-threads).  }
	\label{fig:scalability}
\end{figure}

We also evaluate the scalability of the tested indexes on write operations
and report the results in~\cref{fig:scalability}.
Overall, except for the two concurrent indexes, which do not scale due to lock
contention, all indexes scale well on 224 hyper-threads. This indicates that
their performance is mostly bounded by the work and I/O efficiency of the
underlying algorithms.  \ourlib{} generally achieves the best scalability for
tree construction on 224 hyper-threads (75.6$\times$), much better than the
fastest baseline \cpam{} (50.7$\times$).  Similar trends appear for batch
updates as well.  \pma{} also scales well for single batch updates, since
applying one batch of $1\%$ of the tree size rarely triggers a reconstruction.


\section{Related Work}\label{sec:related}


\myparagraph{Parallel \btree{s}.}  Since the \btree{}~\cite{bayer1972,comer1979}
was introduced in the 1970s, many parallel variants have been proposed in the
literature.
Early parallel \btree{s} in the PRAM model focused on improving the bounds on
construction~\cite{deo1994parallel} and
updates~\cite{higham1994maintaining,lim2019optimal,park2004deleting,braginsky2012lock}.
These approaches are based on pipelining and rely on cycle-based
synchronization, which makes them unsuited for modern hardware.
Goodrich et
al.~\cite{goodrich2025parallel} attempted to adapt the Join-based
framework~\cite{blelloch2016just} to \btree{s};
however, the resulting algorithm exhibits high design complexity, posing significant challenges for practical implementation.
These works are purely theoretical, however. 

Many practical implementations---for
DRAM~\cite{tian2019transforming,wang2018building,lehman1981efficient,xu2023bp},
disk~\cite{bender2005concurrent,roh2014b,srinivasan1991performance}, persistent
memory~\cite{chen2020utree,liu2026ptree,du2025buffered}, and
GPUs~\cite{zhang2022high,yan2019harmonia,daga2012exploiting}---rely on
\emph{locks} to handle node splits/merges.
When processing batch updates, such designs inevitably incur high synchronization overhead due to intense lock contention, which degrades the performance.

Various batch-parallel BSTs have been developed.
Our main baseline, PAM~\cite{sun2018pam}, uses weight-balanced trees.
Later works, Aspen~\cite{dhulipala2019low} and CPAM~\cite{dhulipala2022pac}, extend PAM with enlarged tree nodes to improve I/O performance,
yet preserve the binary layout, partly to support functional updates via path copying.
Unlike these systems, we focus on in-place updates.
Ultimately, their binary structure prevents them from achieving non-trivial I/O bounds.

Several lock-free parallel \btree{s} exist.  The PALM tree~\cite{sewall2011palm}
performs batch updates in the Bulk Synchronous Parallel (BSP)
model~\cite{gerbessiotis1994direct}.  Like \pbtree{}, it splits overflow nodes during
insertion and reinserts underflow nodes.  However, its BSP design targets
communication cost and load balancing, exposing limited parallelism: work within
each node is pinned to a single thread, so underflow reinsertion is sequential.
It also provides no I/O optimization or theoretical guarantees.
Akhremtsev et al.~\cite{akhremtsev2016fast} present a parallel $(a,b)$-tree with provable work and span bounds
for batch updates. However, their analysis assumes constant $a$ and $b$,
and therefore also does not yield I/O bounds. Since their code is not publicly available,
we could not include it in our evaluation.

\myparagraph{Concurrent Search Indexes.} Concurrent \bbplustree{s}, such as
\masstree{}~\cite{mao2012cache} and \bwtree{}~\cite{wang2018building}, are widely studied~\cite{bayer1977concurrency, cha2001olfit, lehman1981efficient, graefe2010survey, oukid2016fptree, wrench2026concurrent,nelson2022bundling,choe2022hybrids} due
to their prevalence in storage systems and databases. Concurrent data structures
target a different setting of simultaneous point updates, and thus rarely
deliver peak performance on batch-parallel workloads. However, applying updates
individually makes single-update latency independent of batch size. On the other
hand, parallel structures (including our \ourtree{}) suit batch-based
applications such as those in graph processing~\cite{dhulipala2019low,wheatman2024byo,pandey2021terrace}.

\section{Conclusion\label{sec:conclusion}}
We introduce \ourtree{}, the first batch-parallel \btree{} to pair provable
guarantees on work, span, and cache complexity with state-of-the-art practical
performance. Its design rests on a single invariant---a repair never touches a
sibling and is always pushed to the parent---one that reduces every overflow and
underflow to one uniform, \overflow{}-based mechanism.  This simplicity is what
makes \ourtree{} scale: repairs stay on the ancestor path, so the
divide-and-conquer structure exposes ample parallelism with little coordination
overhead. As a result, across synthetic, key--value, and temporal-graph
workloads, \ourtree{} achieves the fastest batch updates among the baselines
while remaining competitive on queries. Natural next steps include extending it to
the external- and persistent-memory settings and interleaving concurrent point
queries with batched updates.


%
\section{Acknowledgments}
This work is supported by NSF grants IIS-2227669, TI-2346223, NSF CAREER Awards
CCF-2238358, CCF-2339310, and OAC-2543534, and the Google Research Scholar
Program.


\bibliographystyle{ACM-Reference-Format}
\balance
\bibliography{bib/strings, bib/main}

\iffullversion{
	\clearpage
	\appendix

	\begin{figure}[t]
	\includegraphics[width=0.48\textwidth]{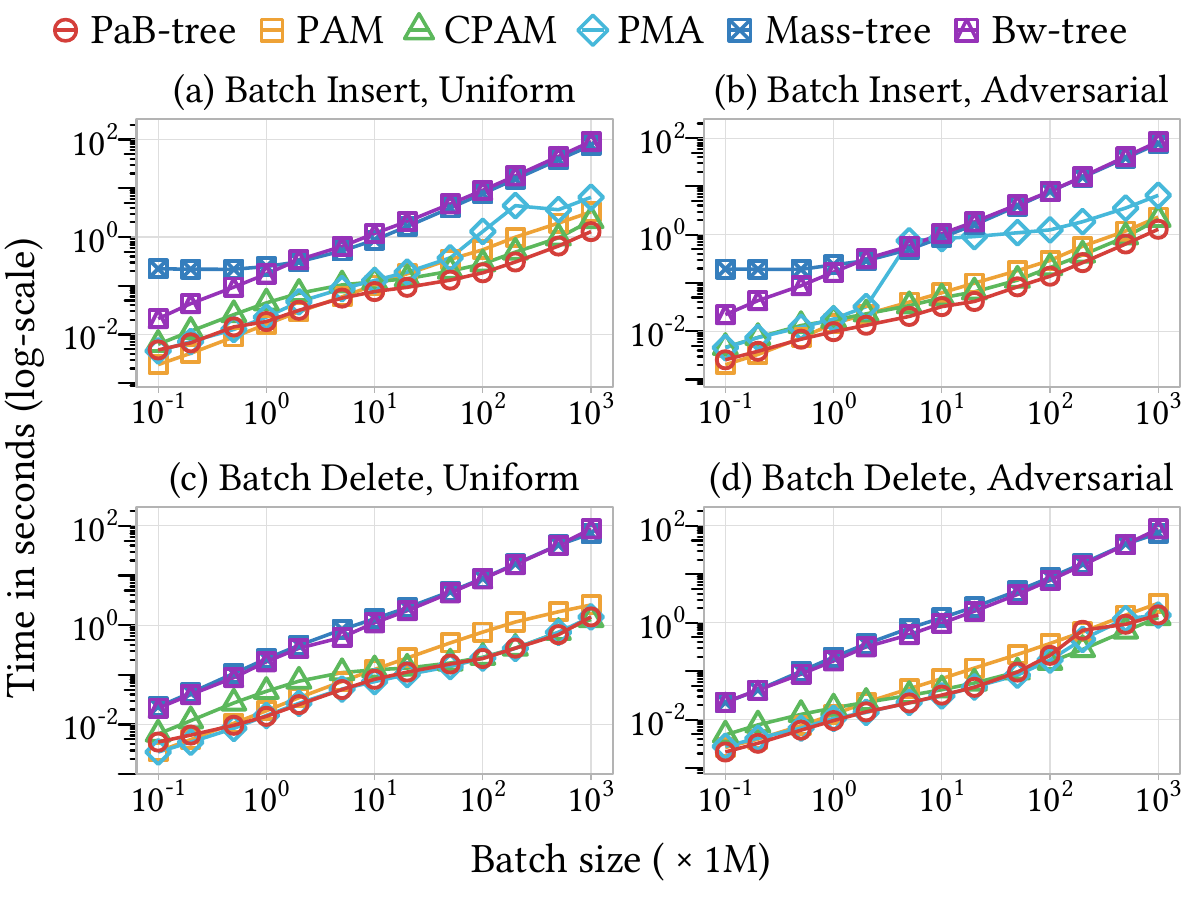}
	\caption{
		\textbf{Running time (in seconds) of single batch update on synthetic datasets. Lower is better.}
		The tree contains $10^9$ entries, and the batch size ranges from $10^5$ to $10^9$.
		The batch insertion is to insert another batch from same distribution, and the batch deletion is to delete a batch of entries from the tree.
		Plots are in log-log scale.
              }
	      \label{fig:batch_updates}
\end{figure}


\begin{figure}[t]
	\includegraphics[width=0.48\textwidth]{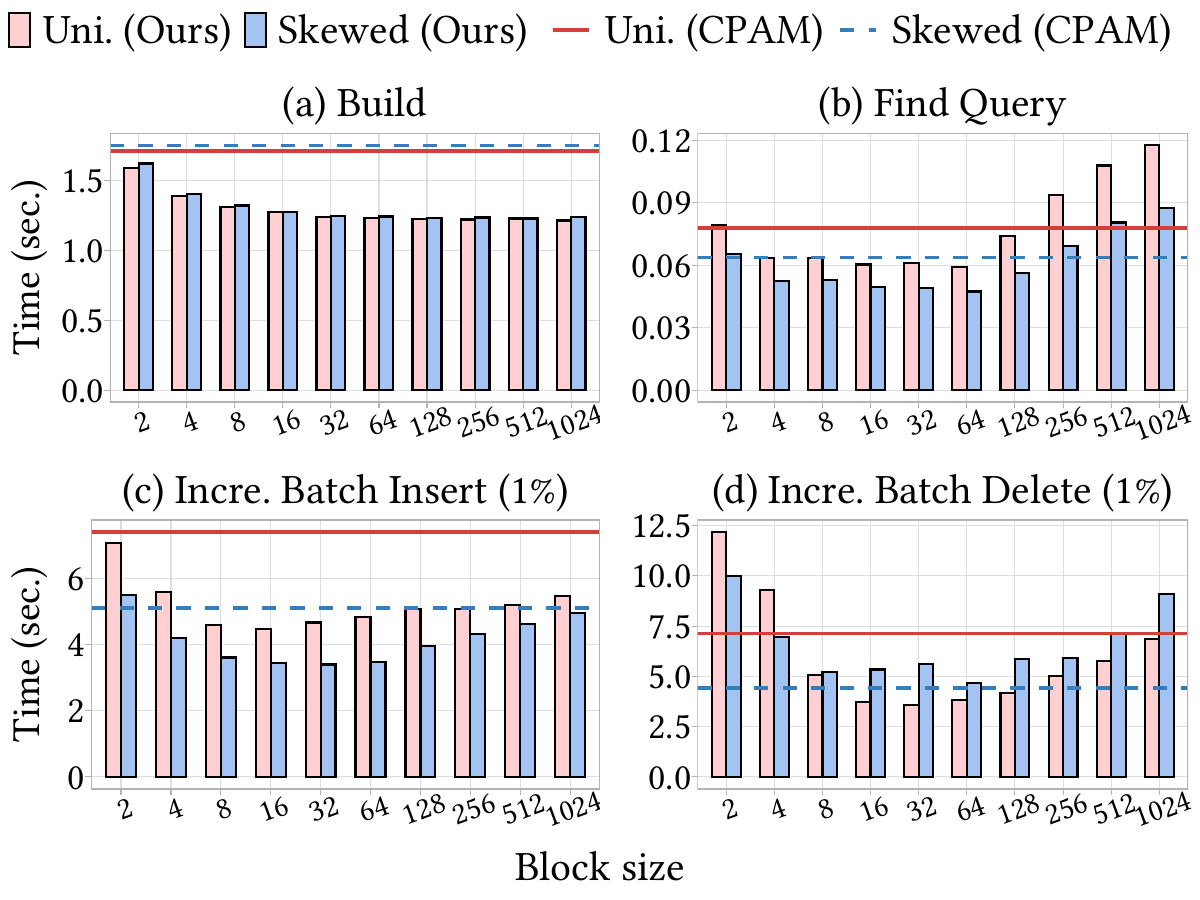}
	\caption{\textbf{Running time (in seconds) of tree construction, incremental batch updates and \find{} queries w.r.t different block sizes in \ourlib{} on synthetic datasets. Lower is better.}
		The time for \cpam{} on the same datasets is marked in lines.
		We measure both the \uniform{} and \ske{} distribution.
	}
	\label{fig:block_size}
\end{figure}

\section{Single Batch Updates Evaluation\label{app:single-batch}}
We measure the time for \ourlib{} and the baselines to apply a single batch of
insertions or deletions on the \uniform{} and \adv{} datasets.  All baselines
behave on \ske{} much as they do on \uniform{}, so we omit it.
We construct a tree with $10^9$ entries, apply one batch to it, and report the
running time.  The batch size ranges from $10^5$ to $10^9$; we exclude smaller
batches because their low computation cost diminishes the practical benefits of parallelism.
We summarize the results in \cref{fig:batch_updates}.

In the incremental setting of \cref{sec:synthetic}, the running time is
dominated by handling invalid subtrees, while a single batch update behaves more
like the static case: for small batches, routing the entries to their leaves
accounts for most of the cost; and for large batches, handling invalid
subtrees poses less overhead than it does in the incremental setting, where the
updates trigger invalid subtrees far more often.  Overall, \ourlib{} is the
fastest or among the fastest under
both distributions.  On small batches, \cpam{} and \pam{} are slightly slower than \ourlib{}
because of their larger tree height.
The gap narrows as the batch size grows, since
applying the batch invalidates larger subtrees, where both \pam{} and \cpam{} can utilize tree rotations to fix them easily, while \ourlib{} must reinsert every entry in the invalid subtree.
The insertion curve of \pma{} shows
several spikes, most likely because the entries exceed the capacity of the
underlying array and force \pma{} to reallocate the memory and copy every
entry.  For batch deletion, \pma{} performs similarly to the others, since a
freshly constructed tree leaves every leaf block nearly full and a single batch
rarely removes enough entries to trigger an underflow.

\section{Evaluation of the Block Size\label{app:block-size}}
We compare the performance of \ourlib{} under different block sizes on synthetic
datasets.  The block size offers a trade-off between tree height and the cost of
searching within leaves.  A smaller block size yields a taller tree, which
incurs additional rounds of partitioning during construction and batch updates,
but it reduces the search cost within each leaf and thereby accelerates queries,
and vice versa.  We measure the performance of \ourlib{} on tree construction,
incremental batch updates, and membership queries for block sizes $B$ ranging from $2$ to
$1024$.  We evaluate both the \uniform{} and \ske{} distributions, compare
against \cpam{} with a block size of $64$, and summarize the results in
\cref{fig:block_size}.

For tree construction, the time decreases as the block size grows and stabilizes
beyond $B=64$, which is about $1.31\times$ faster than with $B=2$ and 1.39$\times$
faster than the \cpam{}.  For membership queries, a smaller block size produces
a taller tree and thus more pointer chasing, increasing query latency; a larger
block size, in turn, demands more work to search within each leaf, also raising
query time and causing \ourlib{} to fall behind \cpam{} once $B \geq 512$.

\begin{table*}[t]
	\small
	\centering
	\setlength\tabcolsep{2pt} 
	\renewcommand{\arraystretch}{0.95} 
\begin{tabular}{c|c|rrrr|rrrr|rrrr|rrrr}
\hline
\multirow{2}[2]{*}{\textbf{Bench.}} & \multirow{2}[2]{*}{\textbf{Trees}} & \multicolumn{4}{c|}{\textbf{Build (1B)}} & \multicolumn{4}{c|}{\textbf{Incre. Batch Insert (10M)}} & \multicolumn{4}{c|}{\textbf{Incre. Batch Delete (10M)}} & \multicolumn{4}{c}{\textbf{Find (10M)}} \bigstrut[t]\\
      &       & \multicolumn{1}{c}{\textbf{CC}} & \multicolumn{1}{c}{\textbf{Inst}} & \multicolumn{1}{c}{\textbf{CRs}} & \multicolumn{1}{c|}{\textbf{CMs}} & \multicolumn{1}{c}{\textbf{CC}} & \multicolumn{1}{c}{\textbf{Inst}} & \multicolumn{1}{c}{\textbf{CRs}} & \multicolumn{1}{c|}{\textbf{CMs}} & \multicolumn{1}{c}{\textbf{CC}} & \multicolumn{1}{c}{\textbf{Inst}} & \multicolumn{1}{c}{\textbf{CRs}} & \multicolumn{1}{c|}{\textbf{CMs}} & \multicolumn{1}{c}{\textbf{CC}} & \multicolumn{1}{c}{\textbf{Inst}} & \multicolumn{1}{c}{\textbf{CRs}} & \multicolumn{1}{c}{\textbf{CMs}} \bigstrut[b]\\
\hline
\multirow{4}[2]{*}{\uniform} & \ourlib{} & \underline{689102} & \underline{427666} & \underline{2757} & \underline{1896} & \underline{2606644} & 1916753 & \underline{7499} & \underline{6732} & \underline{2527648} & \underline{1131125} & \underline{9065} & \underline{8541} & 41462 & 12126 & 229   & 135 \bigstrut[t]\\
      & PAM   & 1309474 & 641143 & 3676  & 2450  & 5010064 & \underline{1769777} & 14481 & 13654 & 7713895 & 1682153 & 28172 & 26765 & 54587 & \underline{9944} & 257   & 154 \\
      & CPAM  & 1141550 & 460966 & 2913  & 2172  & 3970271 & 2557469 & 14894 & 13246 & 3833292 & 2306154 & 13942 & 12550 & 51803 & 12059 & 322   & 200 \\
      & PMA   & 1659997 & 782120 & 3931  & 3137  & 8056747 & 4005441 & 19041 & 16428 & 5071312 & 2026665 & 14687 & 13641 & \underline{22512} & 12389 & \underline{162} & \underline{63} \bigstrut[b]\\
\hline
\multirow{4}[2]{*}{\ske} & \ourlib{} & \underline{723544} & \underline{434176} & \underline{2759} & \underline{1884} & \underline{1898140} & \underline{1342657} & \underline{5414} & \underline{4731} & 2962518 & \underline{1425902} & 8599  & 7727  & 33791 & 11387 & 196   & 106 \bigstrut[t]\\
      & PAM   & 1145419 & 638249 & 3645  & 2366  & 3692366 & 1497891 & 10573 & 9827  & 5449122 & 1475377 & 19262 & 18105 & 45955 & \underline{10145} & 219   & 118 \\
      & CPAM  & 918078 & 459656 & 2854  & 2123  & 2847642 & 1817654 & 9932  & 8844  & \underline{2379974} & 1453233 & \underline{8032} & \underline{7177} & 41816 & 13034 & 269   & 149 \\
      & PMA   & 1623022 & 810203 & 3888  & 3122  & 12012571 & 5086143 & 28184 & 24814 & 6178408 & 2680818 & 15210 & 13757 & \underline{21433} & 13849 & \underline{134} & \underline{50} \bigstrut[b]\\
\hline
\multirow{4}[2]{*}{\adv} & \ourlib{} & \underline{755753} & \underline{427042} & \underline{2700} & \underline{1847} & 1492662 & 1028816 & \underline{2869} & \underline{2303} & 1501640 & 880613 & 2992  & 2297  & 24772 & 11319 & 118   & 71 \bigstrut[t]\\
      & PAM   & 1069354 & 638485 & 3599  & 2340  & 1836388 & 944630 & 5682  & 4908  & 1684620 & 920743 & 4583  & 3908  & 31828 & \underline{10358} & 128   & 76 \\
      & CPAM  & 932701 & 470555 & 2852  & 2105  & \underline{1486898} & \underline{919756} & 4024  & 3407  & \underline{769352} & \underline{451312} & \underline{1668} & \underline{1270} & 30904 & 13207 & 163   & 101 \\
      & PMA   & 1757184 & 767818 & 3711  & 3094  & 18835645 & 7638258 & 43640 & 39373 & 10353164 & 4490375 & 22971 & 20477 & \underline{16384} & 12977 & \underline{86} & \underline{34} \bigstrut[b]\\
\hline
\end{tabular}%

	\vspace{.2em}
	\caption{\textbf{Hardware profiling of the baselines.} We report the
          number of CPU cycles (CC), instructions (Inst), cache references (CRs)
	  and cache misses (CMs).  The detailed setting is shown in \cref{sec:synthetic}.
	  The \find{} query contains 10M queries on a tree
          with 1B entries.  All units are in millions.  The best result in each
          column is underlined.  }
	\label{tab:cache-table}
\end{table*}


For batch insertion, the running time rebounds under both distributions as the block grows. The reason is twofold: a small block raises the tree height and therefore the number of rounds needed to reach the leaves, while a large block demands more work to restore the order of the entries within each leaf. 
However, the \ourlib{} is still faster than the \cpam{} due to a better cache access pattern for large block sizes.
Similar rebounds are observed for batch deletion as well.
However, for batch deletion, the cost of reinsertion dominates the overall cost, especially for the \ske{} distribution.
A larger block size can absorb more entries during deletion, while it also leads to more entries being involved if a reinsertion is triggered, which makes it much slower than the \cpam{} for $B>64$.

Overall, $B=64$ balances update and query performance best, so we adopt it as
the default block size in all other experiments.

\section{Hardware Profiling Details\label{app:hardware}}
\tabref{cache-table} reports the raw counters behind \cref{fig:cache} and
extends the profiling from the \ske{} distribution to all three synthetic
distributions.  For tree construction, \ourlib{} attains both the lowest
instruction count and the fewest cache misses under every distribution, and the
two counters barely change across different distributions, because the construction process is purely comparison-based and thus resistant to different distributions.
On incremental batch insertion under \uniform{}, \pam{} executes
8\% fewer instructions than \ourlib{}, yet incurs 2.03$\times$ the cache misses, which in turn makes it 1.92$\times$ slower.  
\find{} repeats the pattern: \pam{} has the
lowest instruction count of all indexes in all three distributions, but its
large tree height requires fetching more cache lines and leaves it slower.

For batch deletion, \cpam{} is faster than \ourlib{} on \ske{} and \adv{}, by up to
1.95$\times$.  On \ske{}, \ourlib{} still executes the fewest instructions, and
the gap comes entirely from I/O, whereas on \adv{}, \cpam{} wins on both
counters, since its join-based algorithm can easily split out the single tree that is triggered by applying a batch of consecutive keys, while \ourlib{} must
reinsert every underflow subtree.  Once this shortcut disappears under
\uniform{}, \ourlib{} leads on all four counters again.  For \find{} queries, the \pma{}
incurs about 2.1$\times$ fewer cache misses than the \ourlib{} on every distribution.
This gain is confined to queries on different distributions.
Rebuilding overflowed blocks during incremental
insertion costs \pma{} up to 12.6$\times$ the cycles of \ourlib{}, with the gap
widening as the keys in a batch become more local, as in the \adv{} distribution, which makes \pma{} the
slowest index for updates.
Overall, the hardware profiling confirms the importance of cache efficiency in achieving high-performance parallel \bt{s}, which is the main reason that \ourlib{} outperforms the baselines in most of the experiments.

\section{Handling of Small Batches\label{app:small-batch}} 
As shown in \cref{batchdelete}, batch deletion repairs node underflows by
setting aside dropped subtrees and reinserting them level by level.
For very small deletion batches, the reinsertion overhead may dominate the total cost.
Although \cref{fig:osm} shows that reinsertions occur infrequently
in practice, we discuss how to handle small batches here for completeness. 

\begin{figure}[t]
	\includegraphics[width=0.48\textwidth]{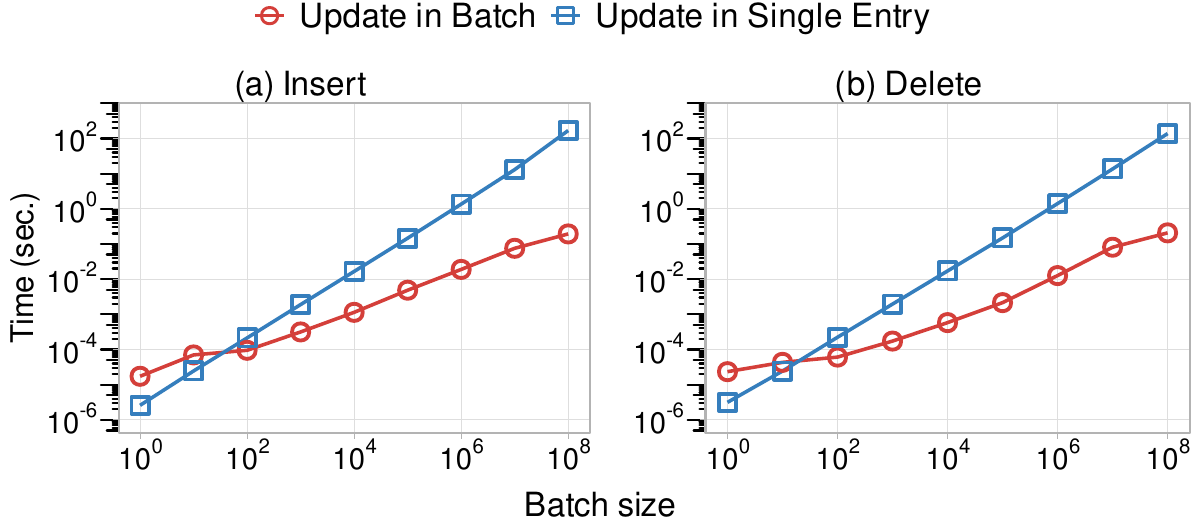}
	\caption{
		\textbf{Running time (in seconds) of batch updates vs. sequential point updates. Lower is better.} We measure the time to apply insertion or deletion batches of different sizes to a tree of $10^9$ entries. Entries in both the tree and the batches are uniformly distributed. Because of the inherent overhead of batching and parallelism, batch updates may be slower than point updates on small batches. However, at a batch size of about 100, the benefits of batching and parallelism start to outweigh this overhead. Accordingly, \ourlib{} can fall back to sequential point updates for batches below this size.
	}
	\label{fig:serial_batch}
\end{figure}

More generally, small batches may diminish the benefits of batch-parallel updates because each thread receives too little work. Existing work on parallel data
structures such as \cpam{} has shown that below a certain batch size,
applying a batch as a sequence of point updates is faster. In any case, the cost of
such small batches is low. We test \ourlib{} with very small batches on
a tree of $10^9$ entries under the \uniform{} distribution and show the results in \cref{fig:serial_batch}. Batch updates
start to outperform point updates at a batch size of about 100, where a batch
takes less than 1 millisecond. Below this cutoff, \ourlib{} can simply fall back
to sequential point updates, which are even faster.

}
\end{document}
\endinput
